\documentclass[twocolumn]{aastex631}
\usepackage[utf8]{inputenc}
\usepackage{amsmath}
\usepackage{amssymb}
\usepackage{xcolor}
\usepackage{natbib}
\usepackage{graphicx}
\usepackage{ulem}
\usepackage{bm}
\usepackage[shortlabels]{enumitem}
\usepackage{appendix}

\shorttitle{KHi turbulence wave damping}
\shortauthors{Hillier et al.}

\begin{document}

\title{Nonlinear wave damping by Kelvin-Helmholtz instability-induced turbulence}
%\author{Andrew Hillier, I\~nigo Arregui \& Takeshi Matsumoto}

\author[0000-0002-0851-5362]{Andrew Hillier}
\affiliation{Department of Mathematics and Statistics, University of Exeter, Exeter, EX4 4QF UK}

\author[0000-0002-7008-7661]{I\~nigo Arregui}
\affiliation{Instituto de Astrof\'{\i}sica de Canarias, V\'{\i}a L\'actea S/N, E-38205 La Laguna, Tenerife, Spain}
\affiliation{Departamento de Astrof\'{\i}sica, Universidad de La Laguna,
E-38206 La Laguna, Tenerife, Spain}

\author{Takeshi Matsumoto}
\affiliation{Department of Physics, Kyoto university, Kitashirakawa Oiwakecho Sakyoku Kyoto, 606-8502, Japan}

%\date{\today}

%\maketitle

\begin{abstract}
    Magnetohydrodynamic kink waves naturally form as a consequence of perturbations to a structured medium, for example transverse oscillations of coronal loops. 
    Linear theory has provided many insights in the evolution of linear oscillations, and results from these models are often applied to infer information about the solar corona from observed wave periods and damping times. 
    However, simulations show that nonlinear kink waves can host the Kelvin-Helmholtz instability (KHi) which subsequently creates turbulence in the loop, dynamics which are beyond linear models.   
    In this paper we investigate the evolution of KHi-induced turbulence on the surface of a flux tube where a non-linear fundamental kink-mode has been excited. We control our numerical experiment so that we induce the KHi without exciting resonant absorption. 
    We find two stages in the KHi turbulence dynamics. In the first stage, we show that the classic model of a KHi turbulent layer growing $\propto t$ is applicable. We adapt this model to make accurate predictions for damping of the oscillation and turbulent heating as a consequence of the KHi dynamics.
    In the second stage, the now dominant turbulent motions are undergoing decay. We find that the classic model of energy decay proportional to $t^{-2}$ approximately holds and provides an accurate prediction of the heating in this phase.
    Our results show that we can develop simple models for the turbulent evolution of a non-linear kink wave, but the damping profiles produced are distinct from those of linear theory that are commonly used to confront theory and observations. 
\end{abstract}

\section{Introduction}

Observations show that the solar atmosphere is filled with highly structured plasma forming loops and threads of material tracing the magnetic field. A number of different phenomena lead to oscillations of these structures in the direction transverse to the field. A few examples are the flare-generated transverse coronal loop oscillations {\citep{aschwanden99,nakariakov99}; the more prevalent small-scale disturbances in active region loops \citep{mcintosh11}; Doppler-shift disturbances in extended regions of the solar corona \citep{tomczyk07}; propagating transverse waves in prominences \citep{lin07,schmieder13}; or the occurrence of transverse waves generated from colliding plasma flows \citep{antolin18}. A common interpretation has been given to these oscillations, in terms of propagating or standing transverse magnetohydrodynamic (MHD) kink waves \cite[see e.g.,][for comprehensive reviews]{ruderman09,goossens11,nakariakov21}.}

The usual paradigm {under which theoretical studies on MHD kink waves have been carried out} is that of waves modelled as {linear perturbations to an initial background state \citep{roberts83,roberts00}. Theoretical models for the damping of kink waves have also focused mainly on the linear regime \citep{goossens92,goossens02a,ruderman02,goossens06}}. However, {as evidenced by the catalogues compiled by \cite{goddard16b} and \cite{NEC2019}, many of the observed kink oscillations have amplitudes that are large and their damping seems to depend on the oscillation amplitude.} For at least some of the observed oscillations, the nonlinearities associated with the perturbations to the system are non-negligible. This can lead to damping of the oscillations through nonlinearities \citep[e.g.][]{CHEN2007, VD2021, ARREGUI2021}.

A well-established result from analytical analysis and numerical simulations is that plasma  motions in a flux tube undergoing a nonlinear kink oscillation can lead to the development of the Kelvin-Helmholtz instability {(KHi)} \citep[see e.g.][]{TERR2008, ANT2014, ANT2015, MAGYAR2016} and the subsequent turbulence the instability can induce \citep[e.g.][]{HILL2019c, AH2020}. 
In this case, the Kelvin-Helmholtz instability is a parasitic instability that grows on the shear-flow that exists on the flanks of the oscillating flux tube. If the instability can grow, it is then able to develop nonlinearities and if there is sufficient energy in the flow it can then develop turbulence \citep{HILL2019b, HILL2019c}. This turbulence can extract energy from an oscillation, either providing a saturation mechanism for the amplitude of the wave for a driven oscillation \citep{AH2020} or lead to damping of an impulsively excited mode. The turbulence excited by the Kelvin-Helmholtz instability is one that is fundamentally different from MHD wave turbulence \citep[e.g.][]{VANBALLEGOOIJEN2011}, where nonlinear MHD waves interact to create a daughter wave of higher frequency. The key difference being that even though both of these mechanisms use the large-scale oscillation as the energy source, in the wave turbulence model the large-scale oscillation is also involved in creating the energy cascade process. For Kelvin-Helmholtz turbulence this is not the case, with the energy cascade being related to the nonlinearities of an instability.

In this paper we perform a detailed analysis of the evolution of the Kelvin-Helmholtz instability on the surface of an oscillating flux tube, identifying how the turbulent dynamics results in two different phases of the evolution of the oscillation amplitude of the tube. We then develop an analytic model for the first stage of the evolution of the flux tube. Here we focus on the large scale response of the oscillating tube to the Kelvin-Helmholtz turbulence, in particular the damping of the oscillations and the heating this creates. 
During the second phase of the dynamics we find that the turbulent energy dominates that of the wave energy, so we develop a model only for this based on classic models of decaying turbulence, using these to predict the heating rate in the latter stage.

\section{Simulations of kink-wave-driven Kelvin-Helmholtz instability and related wave damping}\label{sim_compare}

To build a model of wave damping through Kelvin-Helmholtz turbulence, it is necessary to first identify the fundamental processes that are occurring. To do this we perform a 3D ideal MHD simulation  of a nonlinear kink oscillation that develops the Kelvin-Helmholtz instability. Through this we look to identify what it means for the Kelvin-Helmholtz instability to damp a kink wave, and look at the fundamental processes involved in the damping.

\subsection{Simulation setup}

We perform this ideal MHD simulation of an impulsively excited MHD kink wave using the MHD routines of the (P\underline{I}P) code \citep{HILL2016}. We solve the evolution in 3D {in a Cartesian reference frame} using the non-dimensionalised ideal MHD equations in conservative form, namely:
\begin{align}
\frac{\partial\rho}{\partial t}&+\nabla\cdot(\rho\mathbf{v})=0, \\
\frac{\partial}{\partial t}(\rho\mathbf{v})+\nabla\cdot&\left(\rho\mathbf{v}\mathbf{v}+P\mathbf{I}-{\mathbf{BB}}+\frac{\mathbf{B}^2}{2}\mathbf{I}\right)=0,\\
\frac{\partial}{\partial t}\left( e+\frac{B^2}{2} \right)&+ \nabla\cdot\left[\mathbf{v}(e+P)-(\mathbf{v}\times\mathbf{B})\times\mathbf{B}\right]=0, \\
\frac{\partial \mathbf{B}}{\partial t}&-\nabla \times (\mathbf{v}\times \mathbf{B})=0,\label{ind_eqn}\\
\nabla\cdot\mathbf{B}&= 0,\\
e & \equiv \frac{P}{\gamma-1}+\frac{1}{2}\rho v^2,
\end{align}
which allows us to calculate the evolution of the primitive variables, i.e the density ($\rho$), velocity field {($\mathbf{v}=(v_x, v_y, v_z)^{T}$)}, pressure ($P$) and magnetic field ($\mathbf{B}$), through the evolution of the relevant conserved quantities. Note that in our normalisation we have taken the magnetic permeability of a vacuum inside $\mathbf{B}$ meaning that the local Alfv\'{e}n speed is given by $V_{\rm A}=|\mathbf{B}|/\sqrt{\rho}$.
These equations are solved using a fourth-order central difference approximation for spatial derivatives (calculated on a uniform mesh) with a four-step Runge-Kutta time integration.
We have {non-dimensionalised} the equations using a characteristic coronal density ($\rho_c$), sound speed ($C_{s}$) and lengthscale ($L_c$).
%{\bf Div B cleaning}

Here we do not include any explicit resistive or viscous terms {in the equations}. Dissipation at some level is inherent in the simulation due to the finite grid size and the use of flux limiters to smooth sharp structures. As the code is written in conservative form, any extraction of energy from the flow or magnetic field results in a corresponding increase in internal energy (i.e. heating). This allows us to quantify the magnitude of any heating that occurs in the simulation whilst being able to run calculations in the least viscous, least resistive regime we can achieve for the resolution.

The initial conditions are of a dense tube, aligned with the direction of the magnetic field, which we then perturbed to excite the fundamental kink mode of the system. The initial density profile is given by 
\begin{equation}\label{den_equation}
    \rho(x,y,z)=\rho_{i}+\frac{\rho_{e}-\rho_{i}}{2}\left(1-\tanh\left(2^6\left(\frac{r}{R} -1 \right)\right) \right),
\end{equation}
with $r=\sqrt{x^2+y^2}$, $R$ the radius of the tube, and $\rho_{i}$ and $\rho_{e}$ the internal density of the tube and the external density. We take $R=0.3$, $\rho_e=1$ and $\rho_i=3$. The initial pressure is uniform throughout the domain taking a value $P(x,y,z)=1/\gamma$ (which gives an initial sound speed of $C_s=1$) and a magnetic field of $\mathbf{B}=(B_x,B_y,B_z)^{\rm T}=\sqrt{2/(\gamma\beta)}(0,0,1)^{\rm T}$, with $\beta=0.05$.
The calculations are performed in a domain of $x \in [-L_x,L_x]$, $y \in [0,L_y]$ and $z \in [0,L_z]$ with $L_x=L_y=1$ and $L_z=10$. We use a grid size of $\Delta x=0.005$, $\Delta y=0.0025$ and $\Delta z = 0.1$.
{The anisotropic resolution is chosen to give most resolution to the structures in the $x$-$y$ plane, with greater emphasis on the $y$ direction to help the growth of the initial instability. This anisotropic resolution in the $x$-$y$ plane will result in an anisotropic numerical diffusion of turbulent structures, where the diffusion associated to the larger grid-scale will likely dominate the diffusion of the system as structures are rotated in that plane by the turbulent motions.}
This tube is perturbed with a lateral velocity perturbation to the tube in the form
\begin{equation}
    v_x(x,y,z)=\frac{V_0}{2}\left(1-\tanh\left(2^6\left(\frac{r}{R} -1 \right)\right) \right)\sin\left(\frac{\pi z}{2L_{z}}\right),
\end{equation}
to excite a fundamental kink mode of the system. We use a value of $V_0=0.2$. We also set the $y$- and $z$-components of the velocity to zero {(i.e. $v_y=v_z=0$)}.

We set the boundaries of our domain to be either periodic or possessing symmetries. The boundaries at $x = -L_x$ and $x= L_x$ are set to be periodic. The boundaries at $y=0$ and $y=L_y$ are symmetric boundaries such that the magnetic field is imposed to be in the ($x$, $z$)- plane, as is the flow. This implies that {$v_y=B_y=0$} on the boundary and {$\partial v_x/\partial y=\partial v_z/\partial y=\partial B_x/\partial y=\partial B_z/\partial y=\partial \rho/\partial y=\partial P/\partial y=0$}. The $z$-boundaries are symmetric with the magnetic field penetrating the boundary with the imposed symmetries on the magnetic and flow field being of those of a node ($v_x=v_y=v_z=0$ and {$\partial B_x/\partial z=\partial B_y/\partial z=\partial B_z/\partial z=\partial \rho/\partial z=\partial P/\partial z=0$} at $z=0$) and anti-node ($B_x=B_y=v_z=0$ and {$\partial v_x/\partial z=\partial v_y/\partial z=\partial B_z/\partial z=\partial \rho/\partial z=\partial P/\partial z=0$} at $z=0$ at $z=L_z$).

\subsection{The evolution of the kink wave and their damping through KHi-induced turbulence}\label{sim_results}

Having been subject to its initial kick, as explained in the previous subsection, the tube begins to oscillate. The wave excited is (or at least dominated by) a fundamental kink mode with frequency $\omega_{\rm KINK}=0.53$ {(calculated from linear theory in the long-wavelength limit)}, resulting in the dense tube oscillating back and forth. For our initial conditions we have set a sharp boundary in density between the tube and the external medium (see Equation \ref{den_equation}). Therefore, these oscillations {are not subject to resonant damping, which requires a continuous non-uniform variation of density such that the fundamental kink mode has its frequency in the Alfv\'en  {continuum \citep{VanDoorsselaere2004, goossens06}}. However, as the natural state for these oscillations is for a strong shear flow to develop between the tube and the external medium {\citep{sakurai91,goossens92}}, the boundary of the oscillating tube becomes unstable to the Kelvin-Helmholtz instability {\citep{TERR2008, ANT2014, ANT2015, MAGYAR2016}}.

Figure \ref{Wave_khi} shows the density evolution of the tube cross-section at the wave apex (the $x-y$ plane at $z=L_z$). These snapshots are taken every 3.2 time units to show the evolution of the oscillation at approximately every quarter period. As the oscillations proceed, we can clearly see the development of Kelvin-Helmholtz roll-ups on the top of the tube. After approximately the first quarter period (t=3.2) the instability has started to grow, though the scales associated with this (and with that the area of the cross-section of the tube that has been disturbed by the development of the instability) are small. As the wave continues to oscillate the Kelvin-Helmholtz vortices become larger, and with this the area of the cross section that has been disturbed by the turbulence keeps on increasing. Ultimately, the vast majority of the tube becomes part of the turbulent layer.

\begin{figure*}
    \centering
    \includegraphics[width=18cm]{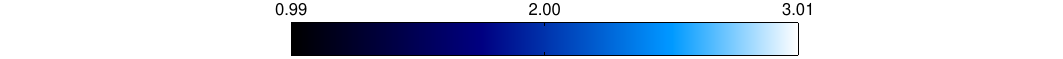}\\
    \includegraphics[width=18cm]{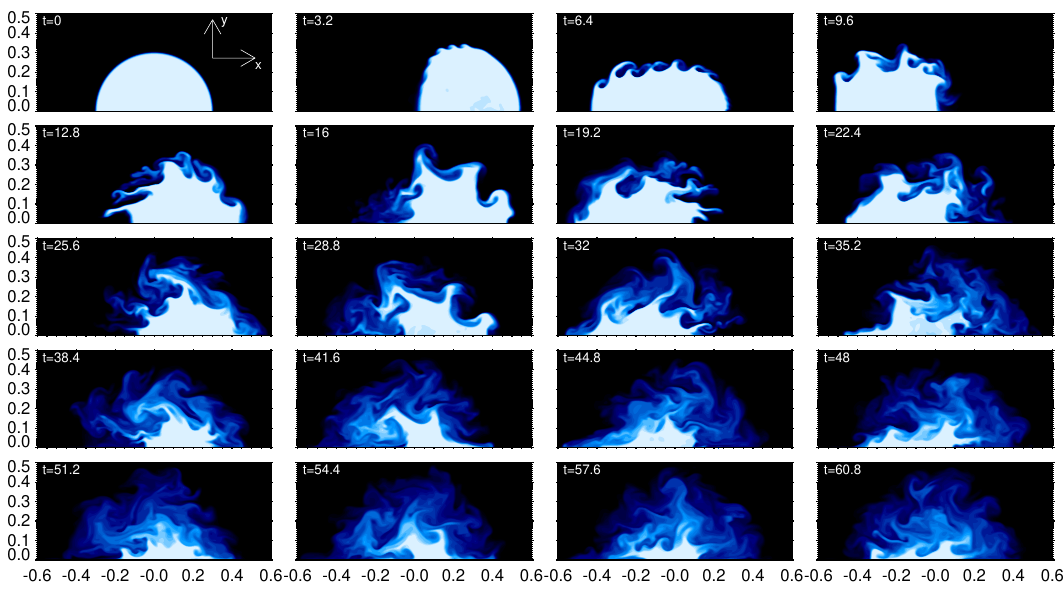}
    \caption{{Contour plots of the density distribution in the ($x$, $y$)-plane taken at $z$= $L_z$ showing the time evolution} of the KHi on the surface of the oscillating flux tube. Time is given in units of the sound crossing time. Colour contours show densities between a lower value of $0.99$ and an upper value of $3.01$.}
    \label{Wave_khi}
\end{figure*}

Looking at this in further detail, Figure \ref{tube_area} shows the temporal evolution of the area of cross section of the tube at the apex (normalised by the initial value) for different density thresholds. We plot the density $\rho\ge 2.7$ (shown by the red line), $\rho\ge 1.89$ (shown by the black line) and $\rho\ge 1.1$ (shown by the blue line). The area shown by the red curve acts as a proxy for the area that has not been disturbed by the Kelvin Helmholtz {induced} turbulence. It is clear that once KHi develops there is a clear evolution of material that is initially part of the kink oscillation of the tube but then joins the mixing layer. Once the undisturbed area of the tube at the apex becomes about 20\% of its initial area, we can see that the decrease in size of the tube core slows. The area of material with $\rho \ge 1.89$ stays approximately constant over time. This is an important result as it connects to the model of \citet{HILL2019c} where for the density contrast used in this calculation the density value of $\rho=1.89$ is predicted to be an area conserving threshold, so is predicted to show no evolution.

\begin{figure}
    \centering
    \includegraphics[width=9cm]{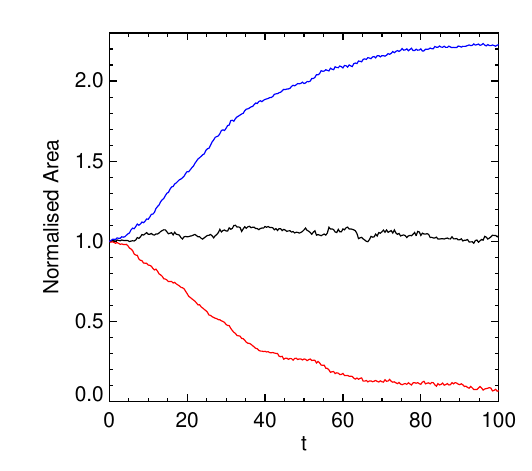}
    \caption{Normalised area of the tube cross section at the apex that has $\rho\ge 2.7$ (red line),  $\rho\ge1.89$ (black line) and $\rho\ge1.1$ (blue line).} 
    \label{tube_area}
\end{figure}

Looking back at Figure \ref{Wave_khi} we can see by eye that the total displacement of the tube, at least as measured by the displacement along the $x$-axis, does not appear to be showing a strong decay with time. We can see this somewhat clearer in Figure \ref{vel_contour} which shows the spatial evolution of $v_x$ at $y=0$ and $z=L_z$ over time. The blue and red colours show the positive and negative velocities respectively. Until a time of $t\approx 44$, the magnitude of the displacement and the magnitude and coherency of the velocity field do not show any drastic change. However, after that time, corresponding to where the decrease in area for the density threshold of $\rho\ge 2.7$ is clearly reduced in Figure \ref{tube_area}, there is a complete change in the dynamics with clear reductions in the lateral displacement of the tube and a velocity field which is less coherent and of smaller magnitude.

\begin{figure*}
    \centering
    \includegraphics[width=16cm]{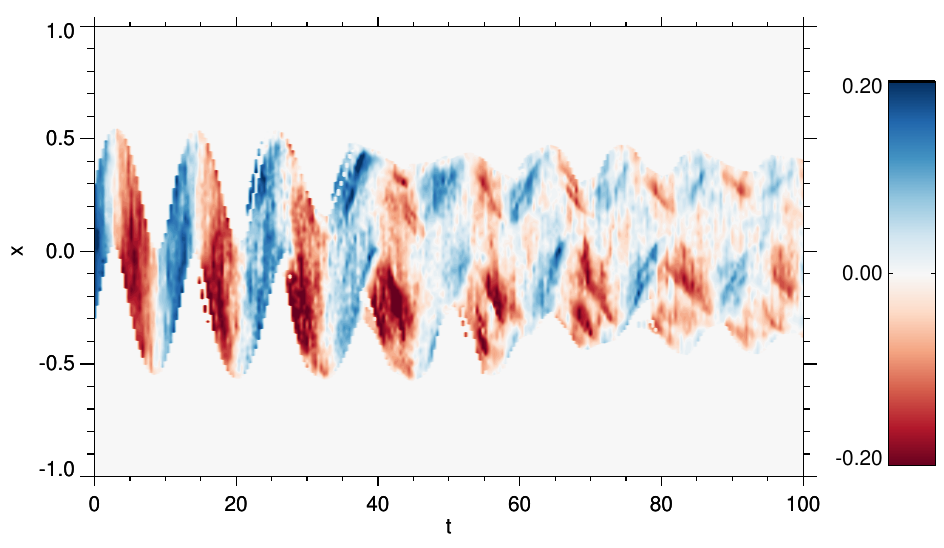}
    \caption{Contour plot of the x velocity in the $x$--$t$ plane at $y=0$ and $z=L_z$. blue (red) colours show positive (negative) values. Colour contour is only given for regions with $\rho\ge1.62$.} 
    \label{vel_contour}
\end{figure*}

The temporal evolution of the velocity field in the plane of the apex provides further details on the evolution from coherent oscillations to turbulent motions. Figure \ref{plane_vel_contour} shows the evolution of $v_x$ in the plane at the apex of the oscillation ($z=L_z$). In this plot we have added two black lines to approximately denote the inner and outer edges of the turbulent layer using the density thresholds of $\rho=2.7$ and $\rho=1.1$ respectively. Though this is more pronounced initially, throughout the evolution the $x$ velocity for the region inside the inner black line (i.e. the core of the tube) is relatively coherent at all times. However, in the turbulent layer (i.e. the region between the two black lines) there is more significant fluctuation around the average motion. This includes local speed-ups in the flow which are a common feature of Kelvin-Helmholtz roll-ups \citep[e.g.][]{HASEGAWA2006} and other turbulent structuring. The coherent component of motions in the turbulent layer clearly do not have to move with the core of the tube, where sometimes they both move in the same direction but sometimes they are moving in opposite directions. 

{We can quantify the difference in the velocity field between the core and the layer by looking at the RMS (Root-Mean-Squared) values of the $x$ velocity and the $y$ derivative of the $x$ velocity in each of the regions. By dividing the former by the latter we can get an approximation of the lengthscale over which the flow is evolving. In the core this value is consistently around 10, implying the average variation of the flow happens on scales larger than the core. However, for the mixing layer this value is $\sim 0.02$, consistent with a flow varying over small length scales.} 

There is one question that Figure \ref{plane_vel_contour} (along with Figures \ref{Wave_khi} and \ref{vel_contour}) elicits related to measuring how the wave damps. We can see that at least for the first few periods the core of the tube undergoes coherent oscillations, but the whole evolution of the tube takes into account the turbulent layer which is growing over time, has more incoherent motions and a coherent component of the motions that moves differently to the core. Therefore, it is necessary to ask: How do we measure the coherent motions of the loop? And are all these measures the same?

\begin{figure*}
    \centering
    \includegraphics[width=18cm]{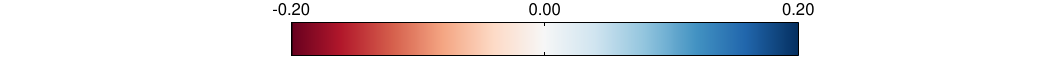}\\
    \includegraphics[width=18cm]{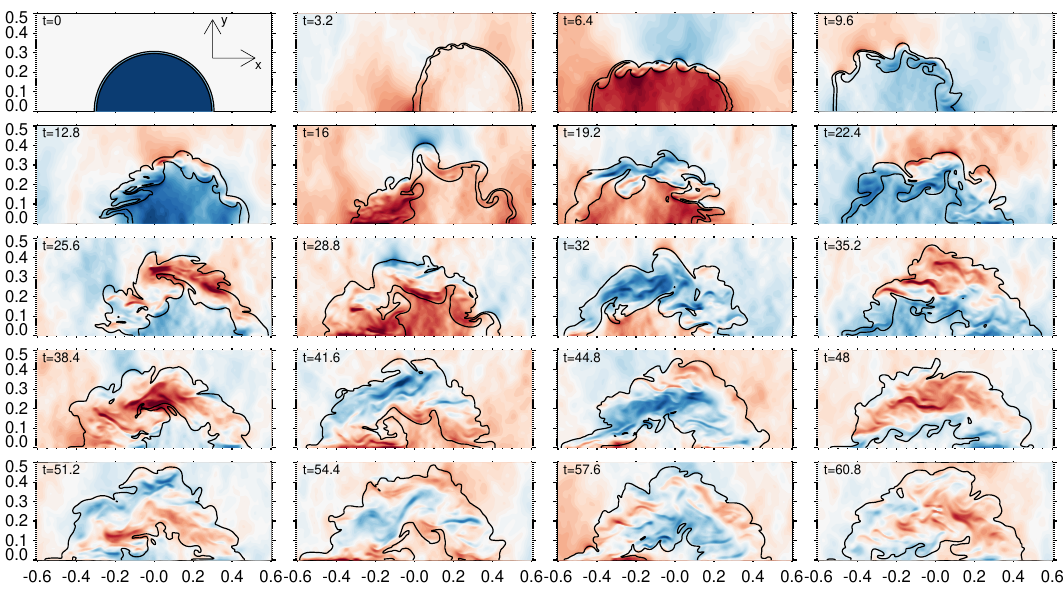}
    \caption{$x$- component of the velocity in the $x$--$y$ plane at $z=L_z$. Colours correspond to same velocities as used in Figure \ref{vel_contour}. Black lines show the $\rho=1.1$ and $\rho=2.7$ transition. snapshots are taken at the same time as those shown in figure \ref{Wave_khi}}
    \label{plane_vel_contour}
\end{figure*}

Clearly the way we measure the bulk motion is determining how we view the change in the oscillation amplitude with time. We can see this very aspect in Figure \ref{damping_profile} where the centre-of-mass displacement in the $x$- direction has been calculated in three slightly different ways resulting in very different displacements measured. For the solid black line, the displacement in the 2D plane at the apex for all the material above a density threshold of $\rho \ge 1.62$ is displayed. The solid red line is a similar calculation, but for a density threshold of $\rho\ge 1.1$ and the blue line for a density threshold of $\rho\ge2.7$. We can see that initially the amplitude of the oscillation, as given by the maximum and minimum displacements, decreases monotonically for all cases. However, for the different density thresholds the rate at which the amplitude reduces (and the form of the envelope giving the damping of the oscillating profile) clearly differs. We can also see that there is a clear period drift for the cases with lower thresholds compared to the $\rho\ge2.7$ curve. After $t\approx 44$ the oscillations are difficult to distinguish in the lower threshold curves, but are clear and consistent in the $\rho\ge2.7$ curve that captures the motion of the core of the tube. 

{Figure \ref{energy_dist} panel (a) shows the kinetic energy distribution at the loop apex (i.e. $z=10$) and panel (b) shows the magnetic energy at the loop foot point ($z=0$). There are three important pieces of information that can be taken from this. Firstly, the density thresholds are reasonably accurate at capturing the extent of the turbulent layer at the apex, but not at the foot point. This can be understood dynamically as the KHi grows at the apex, with flows that transport the density field. These flows, often vortical, will excite waves that travel along the magnetic field and stress the magnetic field at the foot point. However, the fixed boundary means that there are no flows to move the mass. So at the foot point the density thresholds there do not correspond to the width of the turbulent layer (it is more the magnetic mapping from the apex to the foot point that would provide this). Secondly, it can be seen that at this time, the energy is greater in the mixing layer than the core, and the energy fluctuations in the mixing layer are of small scale (as expected from the lengthscale analysis presented above). Thirdly, the area with turbulent kinetic energy seems to be slightly larger than the region of turbulent magnetic energy, but this is balanced by the turbulent magnetic energy generally having larger magnitude.}
 
\begin{figure}
    \centering
    \includegraphics[width=9cm]{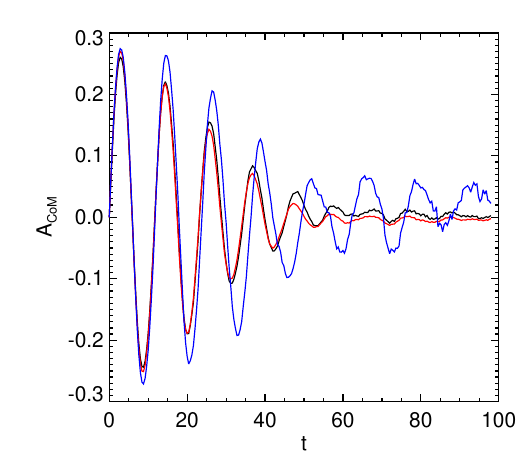}
    \caption{Plot of the evolution of the amplitude of the displacement of the centre-of-mass of the tube in the $x$ direction over time. The three curves are shown for the centre-of-mass calculated for density thresholds of $\rho\ge1.1$ (red), $\rho\ge 1.62$ (black) and $\rho\ge 2.7$ (blue).}
    \label{damping_profile}
\end{figure}

\begin{figure}
    \centering
    \includegraphics[width=9cm]{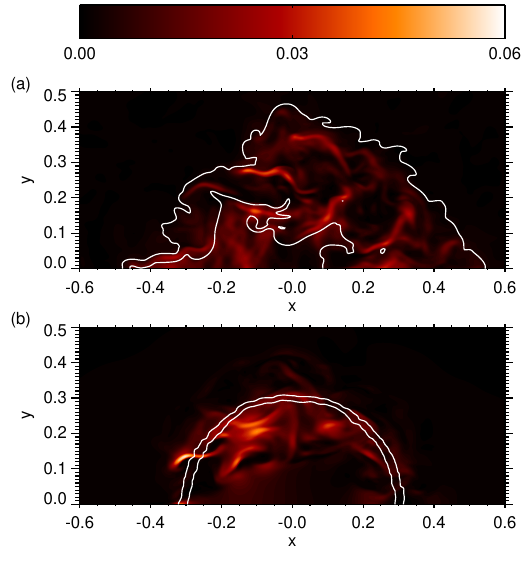}
    \caption{Kinetic energy at the loop apex (panel a) and magnetic energy at the foot point (panel b) at $t=35.2$. White lines show the density contour for $\rho=1.1$ and $\rho=2.7$}
    \label{energy_dist}
\end{figure}

Ultimately this leads us to a fundamental question: How do we quantify damping of a kink oscillation when the loop develops a growing turbulent layer? The standing kink wave is a coherent oscillation of a dense tube aligned with magnetic field. The damping of these oscillations, when considering linear wave theory, should be the same wherever it is measured, not heavily dependent on how the tube is being measured. Clearly something different is happening in this simulation where not only the amplitude but the period of the oscillations measured is a function of the density threshold used to calculate a centre-of-mass amplitude evolution. 

Another key aspect of this simulation is that there are two clear phases of the dynamics. The initial period of damping oscillations connected to the growth of a turbulent layer, followed by the later stage when the tube cross-section is almost completely turbulent and the growth of the layer is almost completely arrested. 
In the next two Sections we will present models for these two phases by developing and combining aspects of the models in \citet{HILL2019c} and \citet{HILLIER2023}, and bench-marking the predictions of the model with key aspects of the simulated evolution.

\section{Modelling the development of a turbulent layer around an oscillating tube}

In this section we develop a formulation to describe the evolution of the kink oscillation into a turbulent tube, i.e. the first stage of the damping as described in the previous section. A model for key aspects of the second stage will be presented in Section \ref{Phase_2}.

The first stage of the damping process happens as the turbulent layer on the boundary grows. This process exchanges momentum from inside and outside of the tube, manifesting as the originally coherent oscillations of the tube developing a more incoherently moving outer layer, with a coherently moving inner region. This can be seen in Figure \ref{damping_profile} where the oscillations of the centre-of-mass for the material with density greater than $1.1$ has an amplitude decaying faster than just the material with density greater than $2.7$.

In this regime, we can expect the velocity jump across the turbulent layer stays relatively constant because the velocity at the centre of the tube (as evidenced by Figure \ref{vel_contour}) remains at approximately the same magnitude until $t\approx 44$. If we consider a simpler flow, i.e. a non-oscillating hydrodynamic shear layer, it is well known that for a constant velocity jump across the layer the thickness ($h$) of the turbulent layer grows as \citep[e.g.][]{winant_browand_1974}
\begin{equation}
    h=\beta_{\rm MIX} \Delta V t,
\end{equation}
where $\beta_{\rm MIX}$ is a constant for a given mixing layer but the particular value will depend on the density contrast of the layer {\citep{BALTZER2020}}. Using a dimensional analysis of the model proposed in \citet{HILL2019c} and developed in \citet{HILLIER2023} we propose that the mixing layer grows as
\begin{equation}
    h=C_1\sqrt{\frac{1}{2}}\frac{(\rho_i\rho_e)^{1/4}}{\sqrt{\rho_i}+\sqrt{\rho_e}} \Delta V t,
\end{equation}
where $C_1$ is a constant of between $\sim 0.1$ and $\sim 1$.
Through comparison with mixing in hydrodynamic shear flows, the value of $C_1$ is expected to be in the range of {$C_1=0.3$ \citep{BALTZER2020} to $0.5$ \citep{BROWN1974} as discussed in \citet{HILLIER2023}}. To determine an appropriate value for the mixing constant $C_1$ for this problem we will look more at the mixing in this section.
By doing this we can then use this model of an expanding shear layer, combined with information on the structure of the shear layer taken from \citet{HILL2019c}, to calculate the rate at which mass and momentum are transferred into or out from the oscillating loop and with that how the oscillation damps.

Before making these comparisons, it is necessary to measure the magnitude of the shear flow across the turbulent layer as this is what drives the Kelvin-Helmholtz instability-induced turbulence.
Figure \ref{shear_flow_profile} shows the shear flow at the surface of the tube measured at $x=0$ and $z=L_z$ at $t=11.6$ (blue line) and $t=18$ (red line). The velocity is reformulated such that the value is positive at $y=0$ for easy comparison. Two {dashed horizontal lines} are added at $v_x=0.15$ and $v_x=-0.05$ to show the approximate values of the flow inside the dense tube and outside the tube. This implies the magnitude of the shear flow is $\Delta V\approx 0.2$. {We can see that locally the shear may appear to be larger than this value, but this is a consequence of the KHi vorticies driving local speed-ups in the flow \citep[e.g.][]{HASEGAWA2006}, which can be seen in Figure \ref{plane_vel_contour}, and not a change in the large-scale velocity shear that feeds the turbulence.}

\begin{figure}
    \centering
    \includegraphics[width=9cm]{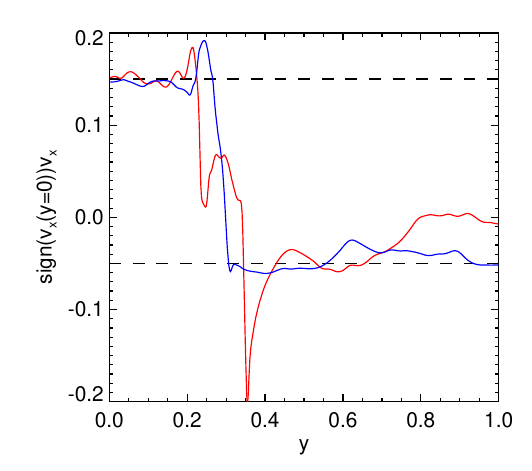}
    \caption{Variation of $v_x$ with $y$ for $t=11.6$ (blue line) and $t=18$ (red line). The sign of $v_x$ is set such that it is positive at $y=0$ for easy comparison.}
    \label{shear_flow_profile}
\end{figure}

Some consideration has to be given to the fact that this is an oscillatory flow and  as such will not have the same magnitude of shear flow at any given time. We hypothesise that as the turbulent motions drive the growth in layer width, the RMS shear flow speed is the appropriate value to use, therefore we redefine $h$ with a factor of $1/\sqrt{2}$, 
\begin{equation}\label{final_h_eqn}
    h=\frac{1}{2} C_1\frac{(\rho_i\rho_e)^{1/4}}{\sqrt{\rho_i}+\sqrt{\rho_e}} \Delta V t.
\end{equation}
Note that this is not strictly necessary as the constant $C_1$ will have to be calculated and the value calculated will just scale up or down based on the numerical constants included in the model. To make the value of $C_1$ as close to unity as possible (i.e. explicitly including as many of the physical processes that determine its value as possible), we include this factor here. 

\subsection{Mass evolution}

To study the evolution of the tube oscillations in Section \ref{sim_results} we looked at the centre of mass motions of the system based on given density thresholds (e.g. Figure \ref{damping_profile}).
However, these different thresholds result in changes in the mass over time, so {a} first important step is to determine how the mass above a given threshold evolves with time.

{As for this model, the actual detailed wave dynamics are not so important for the construction of the model we have decided for simplicity of the model to take a rectangular cross-section of length $2R$ and initial height $H=R\pi/4$ to model the initial cross section of the tube to simplify the calculation of the evolution of the dynamics in the analytical model. The simple justification of this is that this maintains the cross-sectional area, mass, momentum and kinetic energy of the tube whilst making analytic progress easier. Figure \ref{schmatic} gives a diagrammatic representation of the model tube cross-section (panel b) and it evolution as a consequence of the growth of mixing layers (panel c).
With the model geometry and following the self-similar argument, we can predict the evolution of the total mass as}
\begin{equation}
    \frac{d m_{\rho>\rho_T}}{dt}=\frac{d}{dt}(\Delta m_{\rho>\rho_T} h 2R),
\end{equation}
where $m_{\rho>\rho_T}$ is the mass above a given density threshold and $\Delta m_{\rho>\rho_T} $ is the change in mass above that density threshold for a mixing layer of unit width as predicted by the model of \citet{HILL2019c}.

\begin{figure*}
    \centering
    \includegraphics[width=18cm]{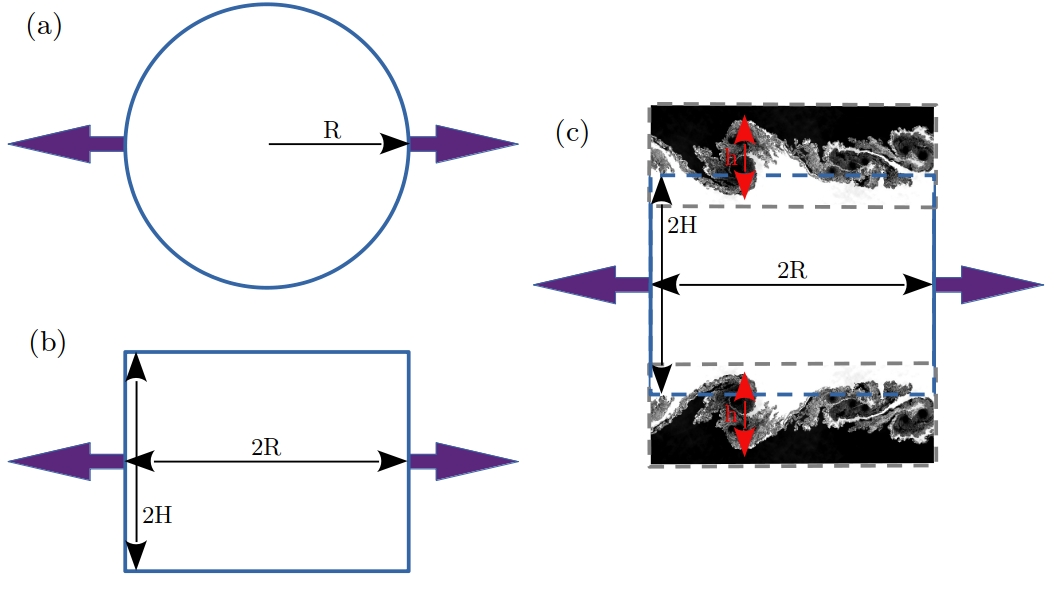}
    \caption{{Schematic to explain the mixing model. Panel (a) shows the cross-section of the tube with radius $R$, the purple arrows show the direction of the oscillations of the tube. Panel (b) shows the geometric manipulation of the tube cross-section used to formulate the analytic model with the width of the now rectangular cross-section kept at $2R$ and the height set as $2H$ with the value of $H$ set to conserve the cross-sectional area. Panel (c) shows how the development of two Kelvin-Helmholtz mixing layers that are growing over time and each have height $h$  (delimited by the red arrows). }}
    \label{schmatic}
\end{figure*}

Based on the model predictions of \citet{HILL2019c}, $\Delta m_{\rho>\rho_T}$ is a constant in time, therefore our model becomes
\begin{equation}\label{mass_eqn}
    \frac{d m_{\rho>\rho_T}}{dt}=\Delta m_{\rho>\rho_T} 2R C_1 \frac{1}{2}\frac{(\rho_i\rho_e)^{1/4}}{\sqrt{\rho_i}+\sqrt{\rho_e}} \Delta V. 
\end{equation}
{Using the model of \citet{HILL2019c}, we can calculate the value of $\Delta m_{\rho>\rho_T}$. This is done by first calculating the initial mass of the dense region before mixing by multiplying the density of the high density region (for our simulations $\rho=3$) by the fraction of a unit length that would have been occupied by this dense phase pre-mixing (i.e. the amount the mixing layer has moved encroached into the the original height of the tube area).} Then we integrate the density component of the model of \citet{HILL2019c} over a unit length from the density threshold value to the maximum density value (in this case $3$). This is plotted in Figure~\ref{mass_values} for different thresholds for the initial density contrast of our simulation. We can see for {a threshold that is both small enough and greater than 1} the mass will be growing in time, but for larger thresholds there is a loss of mass. The value where the mass evolution is roughly neutral, i.e. $\Delta m_{\rho>\rho_T}=0$, is $\rho_{T}=1.62$.  
Equation \ref{mass_eqn} can be integrated to give 
\begin{equation}\label{mass_eqn2}
    m_{\rho>\rho_T}(t)=m_{\rho>\rho_T}(0)+\Delta m_{\rho>\rho_T} R C_1\frac{(\rho_i\rho_e)^{1/4}}{\sqrt{\rho_i}+\sqrt{\rho_e}} \Delta V t. 
\end{equation}

\begin{figure}
    \centering
    \includegraphics[width=9cm]{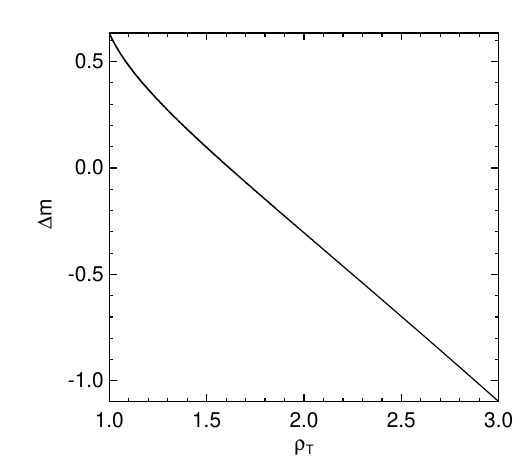}
    \caption{Values of $\Delta m_{\rho>\rho_T}$ for different values of $\rho_T$.}
    \label{mass_values}
\end{figure}

To confirm that Equation \ref{mass_eqn2} gives the expected evolution of the mass, we compare the evolution of the mass for three different density thresholds from the simulation with comparisons to the predicted evolution given by Equation \ref{mass_eqn2}. 
Figure \ref{mixing_mass} shows the evolution of the mass at the apex for three density thresholds: $\rho_T=1.1$ (blue line), $\rho_T=1.62$ (black line) and $\rho_T=2.7$ (red line). The dashed lines are the predicted evolution curves for these three thresholds (during this initial regime of the evolution) using $C_1=0.3$, which was determined by fitting by eye to the $\rho_T=1.1$ lines. The value of $C_1$ found here ($C_1=0.3$) is in the range of the values found in hydrodynamic simulations/experiments \citep{BROWN1974, BALTZER2020}.

Overall the simulated curves follow the predicted linear trend. It is clear that though there is some evolution in the total mass for the density threshold of $\rho_T=1.62$, this is a sufficiently accurate threshold to use to maintain an approximately constant mass in a centre-of-mass velocity calculations. The other two curves show the large variation in mass that can occur when different thresholds are employed. The reason the $\rho_T=1.62$ threshold shows some evolution could be due to errors in applying this Cartesian model for the mixing to this more complex {geometry, with the extreme density thresholds (that capture either almost all or almost none of the mixing layer mass) likely to be less sensitive to this effect as the mapping between the mass associated with the density value in the two geometries being probably more accurate though further work is needed on this.
Other, well-established reasons, unrelated to the mixing process, like the ponderomotive force driving mass accumulation at the loop apex \citep[e.g.][]{TERRADAS2004}, exist as a potential alternative explanation.Though our calculations show that the mass accumulation at the apex by the ponderamotive force can only account for $0.3$\% of the total mass evolution, which is at a level where it can be treated as insignificant.} It is worth noting that the density threshold of $\rho_T=1.89$, which is predicted to be area conserving by the model of \citet{HILL2019}, is approximately area conserving for the duration of the simulation (as evidenced in Figure \ref{tube_area})

\begin{figure}
    \centering
    \includegraphics[width=9cm]{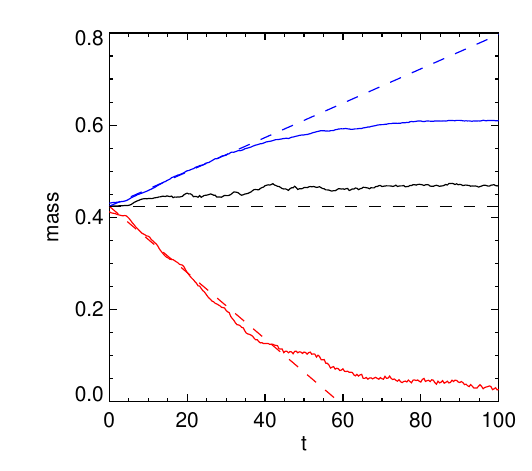}
    \caption{Evolution of the mass of material {at the apex of the tube} found above three different density thresholds. With $\rho_T=1.1$ shown in blue, $\rho_T=1.62$ shown in black, and $\rho_T=2.7$ shown in red. The dashed lines are the predicted evolution value using $C_1=0.3$ with the colours corresponding to the appropriate density threshold.}
    \label{mixing_mass}
\end{figure}

One consequence of having determined the value for $C_1$ is that it allows an upper bound for the time this model holds to be determined.
That is to say we can calculate the time when the mixing layer has become large enough that all of the initial cross-section of the tube is predicted to be engulfed in the turbulent mixing layer. Using the asymmetry of the mixing layer predicted by \citet{HILL2019}, the whole cross-section of the tube should have become turbulent when
\begin{equation}
    \frac{\sqrt{\rho_e}}{\sqrt{\rho_i}+\sqrt{\rho_e}}h(t)=H,
\end{equation}
where as stated previously $H$ is half the initial height used for the rectangular approximation of the tube given as $H=\pi R/4$. This implies that at a time of 
\begin{equation}
    t=\frac{2H}{0.3\Delta V}\frac{(\sqrt{\rho_i}+\sqrt{\rho_e})^2}{\rho_i^{1/4}\rho_e^{3/4}}=44.5,
\end{equation}
the whole tube would have become turbulent and, therefore, the model of a growing turbulent layer is invalid. This predicts that a second stage to the dynamics must have been reached by this time, which is exactly what is seen in Figure \ref{vel_contour}.

\subsection{Momentum evolution}\label{MOM_section}

Following on from the model of the mass evolution, we can develop a similar (though slightly more complex) model for the evolution of the momentum. To do this we split the momentum above a given density threshold into two components: 1) the momentum held in the material still corresponding to the original tube ($M_{\rm core}$), and 2) the momentum injected into the mixing layer ($M_{\rm L}$). This is the same concept as has been used for the modelling of the mass evolution, but with the added complexity that the momentum changes sign as the system undergoes its natural oscillations. The combination of a global oscillation at the kink frequency with the momentum extraction/injection process acts to make the momentum extraction (from the core) or injection (into the layer) oscillatory. Therefore, the forcing in the system driving the momentum change is oscillatory, meaning that it is natural to attempt to approximate this system by a forced linear oscillator.

Firstly, to model the evolution of the momentum in the core, we know that the core of the tube is oscillating at the frequency of the kink wave ($\omega_{\rm KINK}$) and the momentum extraction is driven by the shear-flow dynamics which extract momentum at the same frequency.
Therefore, we need to look at the case where the forcing occurs at the same frequency (i.e. resonant frequency) as the natural oscillations of the system.
This can be stated in equation form as
\begin{equation}\label{MOM_evol_equation}
    \frac{d M_{\rm core}}{d t}=-\omega_{\rm KINK}^2I_{\rm core}-2\dot{F}_{\rm core}\cos(\omega_{\rm KINK} t),
\end{equation}
i.e. a forced linear oscillator where the system is being forced at a resonant frequency.
Here $M_{\rm core}$ is the momentum of the core, we define $I_{\rm core}$ such that $dI_{\rm core}/dt=M_{\rm core}$ and a forcing ($\dot{F}_{\rm core}$) given by
\begin{equation}
    \dot{F}_{\rm core}=2R\overline{M}_0\frac{dh}{dt},
\end{equation}
with $\overline{M}_0$ the initial momentum density given by
\begin{equation}
    \overline{M}_0=\rho_iV_i,\label{core_mom_loss}
\end{equation}
where $V_i$ is the speed of the dense material, in this case $V_i=0.15$ as can be seen in Figure \ref{shear_flow_profile}. {That is to say, Equation \ref{core_mom_loss} gives the rate at which momentum is lost from the core as a consequence of the mixing layer growing at a constant rate. }

Equation \ref{MOM_evol_equation} has the solution
\begin{align}
    I_{\rm core}(t)=&A\sin(\omega_{\rm KINK}t)+B\cos(\omega_{\rm KINK}t)\\&-\frac{\dot{F}_{\rm core}}{\omega_{\rm KINK}}t\sin(\omega_{\rm KINK}t).\nonumber
\end{align}
If we follow our initial assumption that the forcing term is in phase with the oscillations, this implies that $B=-\dot{F}_{\rm core}/\omega_{\rm KINK}^2$, which leads to
\begin{align}
    I_{\rm core}(t)=&\frac{\overline{M}_0}{\omega_{\rm KINK}}\sin(\omega_{\rm KINK}t)-\frac{\dot{F}_{\rm core}}{\omega_{\rm KINK}^2}\cos(\omega_{\rm KINK}t)\nonumber\\&-\frac{\dot{F}_{\rm core}}{\omega_{\rm KINK}}t\sin(\omega_{\rm KINK}t).
\end{align}
This leads to the momentum evolution of the core following
\begin{equation}
    M_{\rm core}=2RH\overline{M}_0 \cos(\omega_{\rm KINK} t) \left(1-\frac{0.3}{{2}H }\frac{\rho_i^{1/4}\rho_e^{3/4}}{(\sqrt{\rho_i}+\sqrt{\rho_e})^2}\Delta V t\right).
\end{equation}

To model the momentum injected into the mixing layer requires more subtlety.
Unlike the mass evolution, the momentum is not a positive definite quantity, it oscillates around zero, so the sign of the momentum injected into the mixing layer changes periodically.
On top of this, the local field lines in the mixing layer have their own characteristic frequency of oscillation (i.e. Alfv\'{e}n frequencies), which can be different from that of the frequency with which momentum is being injected. 

The simple model we put forward to explain how the momentum in the mixing layer develops is based again on a forced linear oscillator, but this time the characteristic frequency of the oscillation of the layer and that of the core are not assumed to be the same.
We consider a single, representative Alfv\'{e}n frequency for the mixing layer (a composite Alfv\'{e}n frequency or kink frequency of the layer) based on the turbulent motions coupling all the different regions of the layer allowing it to develop oscillations at a single representative frequency.
As the model of \citet{HILL2019c} proposes, the representative density of the layer is $\sqrt{\rho_i\rho_e}$. Therefore, the simple estimate of the non-dimensional frequency would be $\omega_{\rm A}=B/(\rho_i\rho_e)^{1/4}$.
However, as the inverse of the square-root of a mean is not the same as the mean of the inverse of a square-root, we calculate the latter from the model of the density distribution across the layer from \citet{HILL2019c} and use this to calculate the approximate Alfv\'{e}n frequency of the mixing layer ($\omega_{\rm A}\approx 0.6$).
Here we model the whole layer as a single forced nonlinear oscillator with momentum $M_{\rm L}$. Mathematically, the evolution of the momentum would be modelled as:
\begin{equation}\label{MOM_evol_equation_layer}
    \frac{d M_{\rm L}}{d t}=-\omega_{\rm A}^2I_{\rm L}-2\dot{F}_L\cos(\omega_{\rm KINK} t),
\end{equation}
where the forcing is occurring at the kink frequency ($\omega_{\rm KINK}$). We define $I_{\rm L}$ such that $dI_{\rm L}/dt=M_{\rm L}$ and $\dot{F}_L$ is the forcing term relating to the rate at which momentum is added into the mixing layer from the external regions. This is given by
\begin{equation}
    \dot{F}_L=\overline{M}C_1\frac{(\rho_i\rho_e)^{1/4}}{\sqrt{\rho_i}+\sqrt{\rho_e}}\Delta V R,
\end{equation}
with $\overline{M}$ as the momentum injection calculated from the model of \citet{HILL2019c} over a region of unit thickness. The precise calculation is given by
\begin{equation}
    \overline{M}=\left(0.15-\frac{\Delta V\sqrt{\rho_e}}{\sqrt{\rho_i}+\sqrt{\rho_e}}\right)\times \int_{y'(\rho_{min})}^{y'(\rho_{max})} \rho dy'.
\end{equation}
The term in the bracket gives the mean velocity of the mixing layer in the rest frame of the simulation. This is then multiplied by the integral of the density between the two density limits of interest across the predicted average density profile of the mixing layer model of \citet{HILL2019c} for a layer with width of unit length. 

Equation \ref{MOM_evol_equation_layer} leads to the well know solution (in the case that the characteristic Alfv\'{e}n frequency of the layer is not the same as the kink frequency of the tube)
\begin{equation}\label{mom_mix_model}
    I_{\rm L}=A \sin(\omega_{\rm A}t)+B\cos(\omega_{\rm A} t)+\frac{2\dot{F}_L}{\omega_{\rm A}^2-\omega_{\rm KINK}^2}\cos(\omega_{\rm KINK} t).
\end{equation}
If we take that at $t=0$ we have $I_{\rm L}=0$ this implies that
\begin{equation}
    B=-\frac{2\dot{F}_L}{\omega_{\rm A}^2-\omega_{\rm KINK}^2}.
\end{equation}
Equation \ref{mom_mix_model} can be differentiated to give the momentum evolution in the mixing layer as
\begin{align}
    M_{\rm L}=&A\omega_{\rm A}\cos(\omega_{\rm A} t)+\frac{2\dot{F}_L\omega_{\rm A}}{\omega_{\rm A}^2-\omega_{\rm KINK}^2}\sin(\omega_{\rm A}t)\nonumber\\&-\frac{2\dot{F}_L\omega_{\rm KINK}}{\omega_{\rm A}^2-\omega_{\rm KINK}^2}\sin(\omega_{\rm KINK} t).
\end{align}
We then have $A=M_{\rm L}(0)/\omega_{\rm A}$ giving
\begin{align}\label{MOM_evolution_pred}
    M_{\rm L}=&M_{\rm L}(0)\cos(\omega_{\rm A} t)+\frac{2\dot{F}_L\omega_{\rm A}}{\omega_{\rm A}^2-\omega_{\rm KINK}^2}\sin(\omega_{\rm A}t)\nonumber\\&-\frac{2\dot{F}_L\omega_{\rm KINK}}{\omega_{\rm A}^2-\omega_{\rm KINK}^2}\sin(\omega_{\rm KINK} t).
\end{align}

\begin{figure}
    \centering
    \includegraphics[width=8cm]{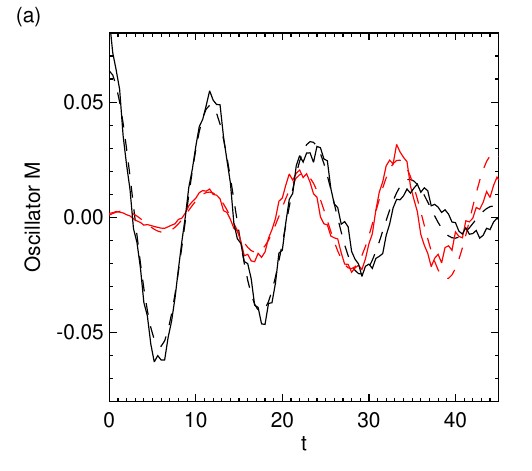}
    \includegraphics[width=8cm]{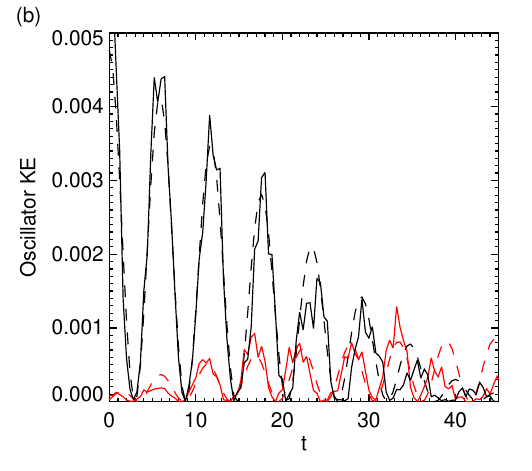}
    \caption{Comparison of the model and the simulation for (a) the momentum evolution and (b) the kinetic energy associated with the bulk motions of the oscillating tube. The black lines show the distribution for material with $\rho>2.7$ for the simulation (solid) and model (dashed). The red lines show the evolution for the quantities where $1.1<\rho<2.7$  for the simulation (solid) and model (dashed). }
    \label{core_mom_damping}
\end{figure}

Figure \ref{core_mom_damping} shows the comparison between the predicted momentum and kinetic evolution of the core of the tube ($M_{\rm core}$) and the mixing layer ($M_{L}$) compared with the simulation results. What is presented here is not a fit, but the solution from an initial value problem compared with the results from the 3D simulations. This model clearly captures the key features of the evolution.

Going further, we can simplify Equation \ref{MOM_evolution_pred} by taking $M_{\rm L}(0)=0$ and looking at the case where $\omega_{\rm A}-\omega_{\rm KINK}$ is small. By setting $\omega_{\rm D}=(\omega_{\rm A}-\omega_{\rm KINK})/2$ and $\omega_{\rm S}=(\omega_{\rm A}+\omega_{\rm KINK})/2$, which means $\omega_{\rm A}=\omega_{\rm S}+\omega_{\rm D}$ and $\omega_{\rm KINK}=\omega_{\rm S}-\omega_{\rm D}$,  we have
\begin{align}
    M_{\rm L}\approx & \frac{\dot{F}(\omega_{\rm S}+\omega_{\rm D})}{2\omega_{\rm S}\omega_{\rm D}}\sin((\omega_{\rm S}+\omega_{\rm D})t)\nonumber \\&-\frac{\dot{F}(\omega_{\rm S}-\omega_{\rm D})}{2\omega_{\rm S}\omega_{\rm D}}\sin((\omega_{\rm S}-\omega_{\rm D})t).
\end{align}
This can be expanded out through double angle formulas to give
\begin{align}
    M_{\rm L}\approx & \frac{\dot{F}}{2\omega_{\rm S}\omega_{\rm D}}\times\\&((\omega_{\rm S}+\omega_{\rm D})(\sin(\omega_{\rm S}t)\cos(\omega_{\rm D}t)+\cos(\omega_{\rm S}t)\sin(\omega_{\rm D}t))\nonumber\\&-(\omega_{\rm S}-\omega_{\rm D})(\sin(\omega_{\rm S}t)\cos(\omega_{\rm D}t)-\cos(\omega_{\rm S}t)\sin(\omega_{\rm D}t)))\nonumber.
\end{align}
Taking that $\omega_{\rm D}\ll\omega_{\rm S}$ this can again be simplified to
\begin{equation}
    M_{\rm L}\approx \frac{\dot{F}}{\omega_{\rm D}}\cos(\omega_{\rm S}t)\sin(\omega_{\rm D}t)).
\end{equation}
This implies we expect our momentum in the mixing layer to oscillate with a frequency of $\omega_{\rm S}$ inside an envelope that evolves as $\sin(\omega_{\rm D}t)$.

Our result above implies that we can refine our prediction for when this initial phase of the dynamics will end. The model we have put forward for this period of the dynamics is that of the oscillations in the core of the tube driving the motions. However, it is clear from panel (b) of Figure \ref{core_mom_damping} (showing the kinetic energy evolution of the core, mixing layer bulk motions and the turbulent fluctuations) that we reach a time where  assuming that the core of the tube holds the majority of the kinetic energy associated with coherent motions no longer holds. At a time of $t\approx 33$ the energy of the coherent motions of the mixing layer becomes greater than that of the energy of the oscillations of the core. So even without including the energy of the turbulent motions we can see that our assumption that the core energetically dominates the dynamics only holds for a limited time. This switch from the core to the layer dominating the energetics might explain why the model is less accurate once we reach a time of $t\approx 40$. 

\subsection{Predicted evolution of amplitude and velocity amplitude of the wave motions}\label{AMP_pred}

With a prediction for both the momentum evolution and the mass evolution above a given density threshold of the oscillation, this can then be turned into an estimate of the centre-of-mass velocity of the oscillation.
{This can be formulated mathematically as
\begin{equation}
    V_{\rm CoM, \rho\ge \rho_{\rm T}}=\frac{\int_{\rho\ge \rho_{\rm T}}\rho v_x da}{\int_{\rho\ge \rho_{\rm T}}\rho da} = \frac{M_{\rho\ge \rho_{\rm T}}}{m_{\rho\ge \rho_{\rm T}}},
\end{equation}}
with $\rho_{\rm T}$ the threshold density used and $a$ the area of the integration. That is the integral of the density weighted velocity divided by the integral of the density gives a centre-of-mass velocity, but this is just the momentum divided by the mass. As both $m_{\rho\ge \rho_{\rm T}}$ and $M_{\rho\ge \rho_{\rm T}}$ have direct predictions as shown in the previous subsections, we can use these to subsequently make a prediction of $V_{\rm CoM, \rho\ge \rho_{\rm T}}$. 

\begin{figure*}
    \centering
    \includegraphics[width=8cm]{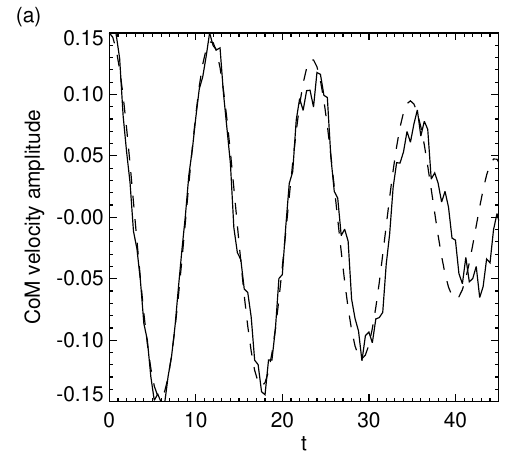}
    \includegraphics[width=8cm]{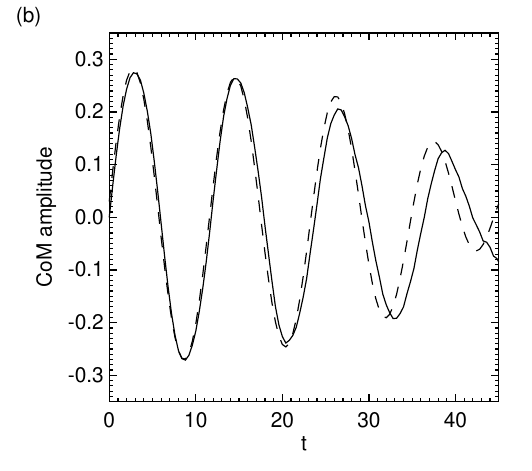}\\
    \includegraphics[width=8cm]{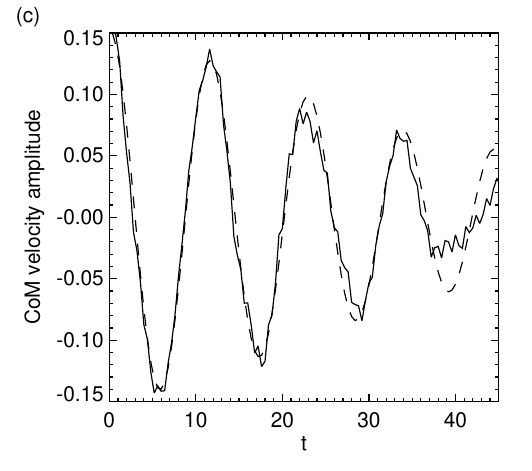}
    \includegraphics[width=8cm]{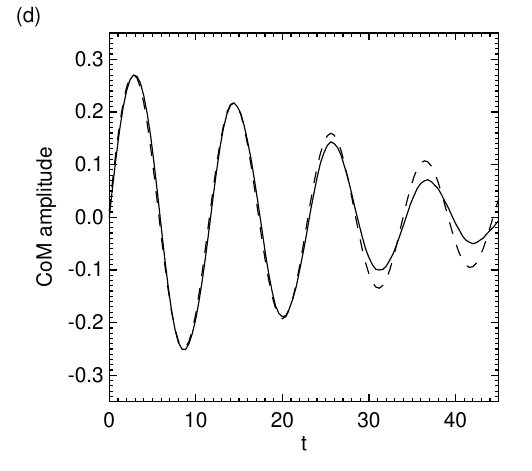}
    \caption{Comparison of the model and the simulation for the centre of mass velocity amplitude (panels (a) and (c)) and the centre-of-mass amplitude of the oscillatory motions (panels (b) and (d)). Panels (a) and (b) show the temporal evolution for the material with $\rho\ge2.7$ and panels (c) and (d) show the temporal evolution for the material with $\rho\ge1.1$. The solid lines are calculated from the simulation and the dashed lines are $V_{\rm CoM, \rho\ge \rho_{\rm T}}$ (panels (a) and (c)) or $A_{\rho\ge \rho_{\rm T}}$ (panels (b) and (d)).}
    \label{vel_and_amp_damping}
\end{figure*}

Panels (a) and (c) of Figure \ref{vel_and_amp_damping} show the temporal evolution of centre-of-mass velocity for just the core of the tube at its apex (panel a) and when the mixing layer is also included (panel c). Unsurprisingly, given the accuracy of the momentum and mass evolution up until a time of $t\approx 40$ these panels show the model is a good representation of the simulation results.

A measure of the evolution of the velocity of the loop apex over time can be used to make a prediction for the evolution of the position of the loop apex (i.e. the wave amplitude) over time. 
Then by integrating $V_{\rm CoM, \rho\ge \rho_{\rm T}}$ over time, we can make a subsequent prediction for the amplitude evolution ($A_{\rho\ge \rho_{\rm T}}$) {of
\begin{align}
    A_{\rho\ge \rho_{\rm T}}=&\int_{0}^{t}\frac{\int_{\rho\ge \rho_{\rm T}}\rho v_x da}{\int_{\rho\ge \rho_{\rm T}}\rho da} dt +A_{\rho\ge \rho_{\rm T}}(0)\nonumber\\=& \int_{0}^{t}\frac{M_{\rho\ge \rho_{\rm T}}}{m_{\rho\ge \rho_{\rm T}}} dt+A_{\rho\ge \rho_{\rm T}}(0).
\end{align}}

Due to the nonlinear system we are studying, though dimensionally the same, the $A_{\rho\ge \rho_{\rm T}}$ defined above is different from the centre-of-mass amplitude which is defined {as
\begin{equation}
    A_{\rm CoM, \rho\ge \rho_{\rm T}}=\frac{\int_{\rho\ge \rho_{\rm T}}\rho x da}{\int_{\rho\ge \rho_{\rm T}}\rho da}.
\end{equation}}
This difference in formulation is very important to consider when comparing results for a nonlinear system, as differing formulations will lead to some differences in results. This is seen in panels (b) and (d) of Figure \ref{vel_and_amp_damping}, where for looking only at the core of the tube (with its simpler evolution) the model prediction for the amplitude evolution ($A_{\rho\ge \rho_{\rm T}}$) is a reasonable representation of the centre-of-mass amplitude from the simulation ($ A_{\rm CoM, \rho\ge \rho_{\rm T}}$). However, in panel (d) there is a much greater decrease in the centre-of-mass amplitude (not seen in the centre-of-mass velocity) compared to the model.

\subsection{Bounding the heating rate in the first phase of the dynamics}\label{phase_1_heat}

An important question, especially in terms of how the results of this paper may impact our understanding of wave heating of the solar corona, is: can we predict the heating rates in the simulation from the model of Kelvin-Helmholtz turbulence. As we have performed an ideal MHD simulation, we do not have any explicit heating terms from which to calculate this, but the numerical dissipation will (due to the use of a conservative scheme) result in energy lost from the kinetic and magnetic energies being added to the internal energy. Combined with the choice of boundary conditions which mean energy cannot leave the simulation domain this gives us some measure of the dissipative heating from the turbulence. The black line in Figure \ref{heating} shows the increase from the initial value of the total internal energy over time. {At the end} of this phase (around $t\sim 40$) we find a total internal energy increase in the simulation volume that is $\approx 0.1$\% of the initial internal energy of the simulation domain.

\begin{figure}
    \centering
    \includegraphics[width=9cm]{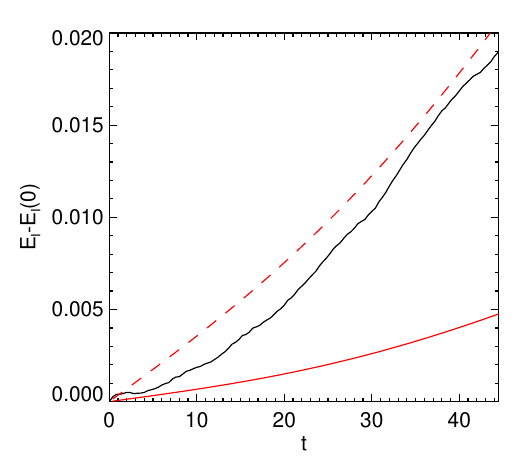}
    \caption{Change in internal energy (solid black line) with predicted slope based on the model of turbulent heating as the Reynolds number tends to infinity (solid red line) and the predicted upper limit of the heating rate (dashed red line).}
    \label{heating}
\end{figure}

To make a prediction of the rate at which the internal energy increases (i.e. a heating rate), first we need to make an estimate {of} how the magnitude of the turbulent energy (both in the velocity and the magnetic fields) held in the mixing layer evolves over time. To do this, we need to account for all the energy that has not been previously accounted for in Section \ref{MOM_section}, i.e. energy not in the bulk motions of the layer. 

This energy can be distributed into three categories: the turbulent energy, the energy in the variation of the mean and the energy lost by the forcing of the mixing layer not being resonant with the layer's natural frequency. The first of these is simple to understand, it is the energy that is predicted to be in turbulent fluctuations by the model of \citet{HILL2019c}. The total energy of this component is calculated as \citep{HILL2019c}
\begin{equation}
    E_{\rm turb}=\frac{1}{16}\frac{\rho_i\rho_e}{(\sqrt{\rho_i}+\sqrt{\rho_e})^2}  \Delta V^2 2R h(t){L_z},
\end{equation}
where a factor of $1/2$ has been introduced as a consequence of {integrating along the loop assuming the velocity shear depends on $z$ as $\sin(\pi z/(2L_z))$.}
The second of these is not seen as turbulent energy by \citet{HILL2019c}, but {has} to do with the mean flow having a distribution across the mixing layer. This enters the calculations here as the mixing layer is no longer moving with the shear driving but is drifting out of phase. As such it is not clear how much of the energy we expect to be in the mean variation is kept there and how much might be released for further turbulence. The total energy of this component is calculated as
\begin{equation}
    E_{\rm mean dist}\approx E_{\rm turb}.
\end{equation}
Finally, we have the energy that is lost by forcing the mixing layer at a frequency different to its natural oscillatory frequency. This is just the difference between the energy of the bulk motions of the mixing layer when they are forced at the kink frequency and forced at the natural frequency of the layer. This difference is more pronounced for larger density contrasts highlighting a greater potential for heating in those situations. As this calculation would already take into account the cross-sectional area, this would be multiplied by $L_z$ to give a total energy $E_{\rm force}$. 

The total possible energy in the turbulence in the mixing layer can then be calculated by summing these three: i.e. $E_{\rm tot}=E_{\rm turb}+E_{\rm mean dist}+E_{\rm force}$. The evolution of $E_{\rm tot}$ is shown with the dashed red line in Figure \ref{heating}. This can be used as an expected upper limit from turbulent heating in the simulation.

We can add more nuance to these arguments. We should expect a time delay between energy being brought into the mixing layer and its dissipation after a turbulent cascade. To model this, we can use the approximate self-similarity of the turbulent energy cascade which implies that the velocity fluctuations at a given length scale ($l$) scale as $\propto (l/l_0)^{1/3}$ where $l_0$ is the largest scale of the turbulent cascade or integral length scale\citep{K41}. Assuming that the turbulent cascade involves the standard nonlinear process where energy is passed from one scale to the scale of half that size and that the turbulence cascades to infinitesimal scales (the Reynolds and magnetic Reynolds number are tending to infinity) leads to the following estimate of the dissipation time $\tau_{\rm DISS}$ \citep[e.g.][]{ONSAGER1949}
\begin{equation}
    \tau_{\rm DISS}=\sum_{i=0}^{\infty}\frac{l_i}{v_i} =\frac{l_0}{v_0}\left(1+ \frac{1}{2^{2/3}}+ \frac{1}{2^{4/3}} +... \right)\approx 2.70 \frac{l_0}{v_0},
\end{equation}
where the subscript denotes the level in the cascade and $\tau_{\rm EDDY}={l_0}/{v_0}$ is the large-scale vortex turnover time. This leads us to estimate that at any given time the energy that has been brought into the mixing layer and has formed part of the turbulent fluctuations, $\tau_{\rm EDDY}/\tau_{\rm DISS}=1/2.7$ of the energy will have been dissipated leaving $1.7/2.7$ as part of the turbulence.

This leads to an estimation of a lower bound for the heating rate of $(E_{\rm turb}+E_{\rm force})/2.7$. This is shown with the red solid line in Figure \ref{heating}. Overall the two bounds we derive do provide bounds for the growth of the internal energy of the simulation. The fact this is closest to the upper bound may imply that the turbulent energy is very efficiently dissipated (in much less than $2.7$ turnover times) or more likely as the energy from the initial kick is not all trapped in the oscillation some extra heating is occurring.

\section{Modelling energy dissipation in the second phase of the dynamics}\label{Phase_2}

Having developed a model to explain the evolution of the first phase, we now turn our attention to the second phase of the dynamics. The key characteristic of this regime that we use to guide the model is that the energy of the dynamics is dominated by the turbulent component (see Figure \ref{turb_energy} after $t\approx 40$). It is also important to note that as no further energy is input into the system and without an energy source the turbulence will decay over time (again see Figure \ref{turb_energy} after $t\approx 40$). Therefore, we focus on developing a model for the decay of the turbulence. 

\begin{figure}
    \centering
    \includegraphics[width=9cm]{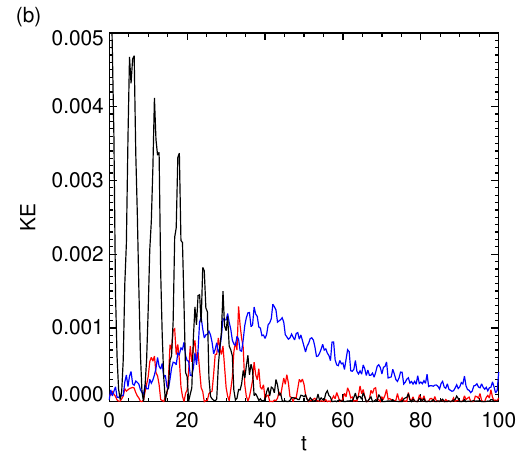}
    \caption{Kinetic energy of the x component of the velocity at the apex of the oscillation for the core of the tube (black line), the coherent motions of the mixing layer {calculated from the mean density and density weighted velocity of the layer} (red line), with the energy of everything left, i.e. the incoherent, turbulent motions (blue line).}
    \label{turb_energy}
\end{figure}

To model the decay of the turbulence, we will no longer consider the energy of the oscillations and restrict our arguments solely to the evolution of the turbulence. This simplification makes the model we propose for understanding the rate at which the turbulence decays to be that of simple decaying turbulence, and independent of the oscillatory dynamics.
The model we use is based on those first proposed by \citet{Taylor1935} and \citet{K41} where the turbulent transport takes the turbulent energy held at large scales to smaller scales until it is dissipated. This can be understood simply by the nonlinear turbulent transport through spatial scales, which can be approximated by:
\begin{equation}\label{turb_eq_1}
    \frac{d E_{\rm turb}}{dt}=-\mathbf{v_{\rm turb}}\cdot\nabla E_{\rm turb},
\end{equation}
where $E_{\rm turb}$ is the energy held in the turbulent fluctuations of the velocity and magnetic field and $\mathbf{v_{\rm turb}}$ are the turbulent motions that transport the energy to smaller scales.

The RHS of Equation \ref{turb_eq_1} can be approximated by using the root-mean-square (RMS) of the turbulent motions ($v_{\rm RMS}$) and the largest lengthscale of the turbulence (also known as the integral lengthscale) and further by connecting to the energy of the turbulent fluctuations to get \citep[e.g.][]{K41, ONSAGER1949}
\begin{equation}\label{turb_eq_2}
    \frac{d E_{\rm turb}}{dt}\approx -C\frac{v_{\rm RMS}}{L(t)}E_{\rm turb}\approx-\frac{C\sqrt{2}}{L(t)A(t)\sqrt{\rho_i\rho_e}}E_{\rm turb}^{3/2},
\end{equation}
where $v_{\rm RMS}\approx\sqrt{2E_{\rm turb}/(A(t)\sqrt{\rho_i\rho_e}})$ \citep[following][]{HILL2019c} is an estimate of the connection between the root-mean-squared velocity of the turbulent motions and the turbulent energy with $A(t)$ the area of the turbulent region and $L(t)$ the integral length-scale of the turbulent motions.
Figure \ref{tube_area} shows that at late times the area of the turbulent region is approximately constant implying that $A(t)$ is constant (which we denote as $A_0$). If the area is not changing then we can also assume that the integral lengthscale $L(t)$ is also constant, which we denote as $l_0$ following Section \ref{phase_1_heat}

Separating the variables in Equation \ref{turb_eq_2} and then integrating leads to the following integral equation{:
\begin{equation}
    \int_{E(t_0)}^{E(t)}E^{-3/2}dE=-\frac{C}{l_0}\sqrt{\frac{2}{A_0\sqrt{\rho_e\rho_i}}}\int_{t_0}^t  dt.
\end{equation}
Therefore,
\begin{equation}\label{turb_energy_eqn}
    E_{\rm turb}=\frac{1}{(\alpha(t-t_0)+E(t_0)^{-1/2})^2},
\end{equation}
with
\begin{equation}
    \alpha= \frac{C}{l_0}\sqrt{\frac{2}{A_0\sqrt{\rho_e\rho_i}}}.
\end{equation}
Based} on the area of the mixing layer shown in Figure \ref{Wave_khi} we take $A_0$ to be an annulus with outer radius of $0.45$ and inner radius of $0.15$. This leads to a lengthscale $l_0$ which we take to be the {difference} between the inner and outer radius of $0.3$. For dynamical consistency we take that the unknown constant $C$ has the {value $D/2.7$ where the $2.7$ comes from the cascade time for Kolmogorov turbulence (see Section \ref{phase_1_heat}), i.e. $C=D/2.7=0.3/2.7$. This then gives 
\begin{equation}
    \alpha= \frac{D}{2.7l_0}\sqrt{\frac{1}{2A_0\sqrt{\rho_e\rho_i}}}.
\end{equation}}

The simple implication from this model is that the turbulent energy should decay at late times $\propto t^{-2}$. This is a classic result for decaying turbulence with a fixed length-scale of the large scale motions \citep[e.g.][]{Taylor1935, Oberlack2002, Sagaut2018}. Figure \ref{turb_decay} shows the comparison between the theoretically predicted decay and that shown for the kinetic energy in the simulation at the apex of the tube {using a value of $D=0.6$, which is twice the value of the constant $C_1$ found in the first stage of the dynamics}. Note that $\alpha$ has been multiplied by a factor of $\sqrt{4/3}$ because we predict that the total kinetic energy will be $4/3$ greater than the kinetic energy in the $x$-direction. This can be inferred from the model of \citet{HILL2019c} due to the approximate equipartition between the energy found in the component of the flow varying about the mean layer velocity for a shear flow and the turbulent component of the flow. Overall the match shown in Figure \ref{turb_decay} between the model and simulation results is good, though this only over a relatively short time range making stronger statements difficult.
\begin{figure}
    \centering
    \includegraphics[width=9cm]{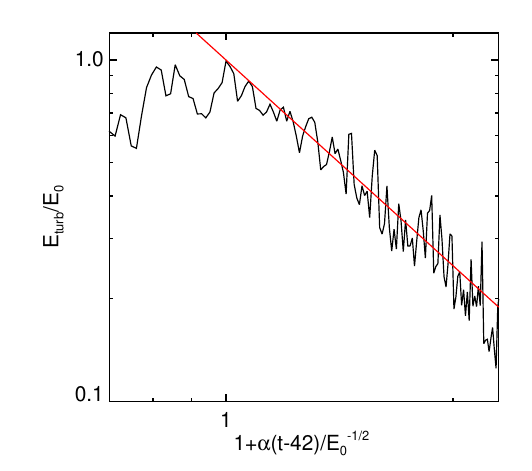}
    \caption{Plot of the normalised turbulent energy against modified time. The black line shows the results of the simulation and the red line the predicted decay from the model.}
    \label{turb_decay}
\end{figure}

\begin{figure}
    \centering
    \includegraphics[width=9cm]{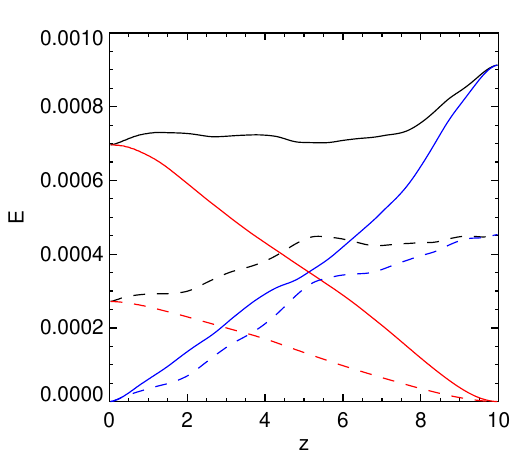}
    \caption{{Plot of the normalised turbulent energy in the $x$ and $y$ components of the magnetic field (red line) and velocity field (blue line) and their total (black line) against z calculated in a tube aligned with $z$ with radius $0.45$. The solid lines are for $t=60$ and the dashed lines for $t=80$.}}
    \label{turb_z_dist}
\end{figure}

If the energy of the turbulence is decaying, the reason for this is that the energy held in the turbulent fluctuations is being dissipated. Therefore, we can expect the decay of the turbulence to directly connect to heating. If the turbulent energy goes down, this should directly correspond to the same level of increase in thermal energy. {To make progress, we assume the turbulent energy (taking the turbulent energy to be the sum of turbulent kinetic energy and turbulent magnetic energy) decay is uniform along the flux tube. Figure \ref{turb_z_dist} shows the distributions of the turbulent kinetic and magnetic energies along the tube at two times. From this we can see that, roughly speaking, the magnitude of the total turbulent energy is approximately constant along the tube at both times implying an approximately constant decay along the tube. 
Making this assumption means we can} integrate $E(t_0)-E_{\rm turb}$ over the length of the tube and this gives the amount of energy dissipated. To take into account the velocity (and with that magnetic field) fluctuations in the $y$-direction we multiply $E(t_0)-E_{\rm turb}$ by $4/3$ and again take the same factor of $\sqrt{4/3}$ in $\alpha$. The comparison between the relative increase in internal energy of the model and the simulation are shown in Figure \ref{turb_heating_2} with the black line giving the simulation results and the red line that of the model (the model is only shown after a time of $t=42$ for which it was derived). The model clearly provides a reasonable representation of the evolution implying that the key physical concepts that are needed to understand the increase in thermal energy of the simulation have been included in the model.

\begin{figure}
    \centering
    \includegraphics[width=9cm]{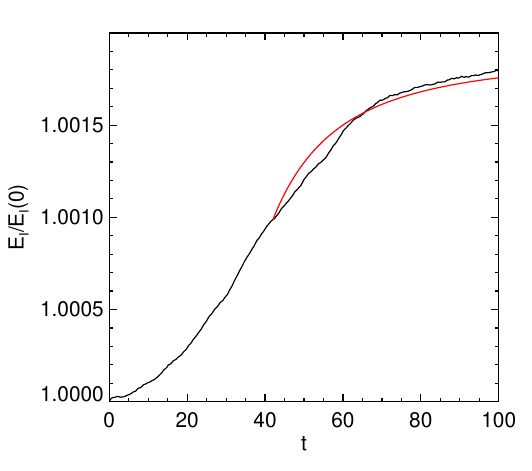}
    \caption{Evolution of total thermal energy over time (black line). Model of the expected increase in thermal energy as a consequence of heating due to decay of turbulence is shown with the red line. {Note that the red line is only plotted for $t\ge 42$.} }
    \label{turb_heating_2}
\end{figure}

\section{Summary and Discussion}

In this paper we have investigated the dynamics of a nonlinear kink wave excited in a flux tube. The numerical study was designed to capture the fundamentals of the system we are interested, i.e. kink oscillations of coronal loops, whilst maintaining sufficient simplicity and conserving energy in the domain to make analysis straight-forward. The oscillating tube develops the Kelvin-Helmholtz instability on its surface, developing a turbulent layer. It is the growth of this turbulent layer that characterises the first stage of the dynamics in the simulation. Once the layer becomes sufficiently large, and it became the dominant source of kinetic energy, the dynamics transition to a second stage characterised by the decay of the turbulent energy in an annulus around the much reduced core of the loop. We presented simplified analytic models to explain key aspects of these two stages.

The first stage of the dynamics is characterised by the growth of a turbulent layer that grows until most of the tube cross-section is turbulent. To model this evolution, we combined together the mixing models of \citet{HILL2019c} and \citet{HILLIER2023} to create new models for how the mass and momentum (and as a consequence energy, wave velocity, wave amplitude and heating) evolve over time.
This model made some clear predictions about when this stage has to end. As this transition is linearly proportional to the magnitude of the shear flow, we would naturally expect that this stage finishes sooner for highly nonlinear waves. 

A fundamental aspect of the model for this stage of the evolution is the self-similar evolution of the turbulent layer. This model was developed for steady hydrodynamic shear flows, e.g. \citet[][]{winant_browand_1974}. However, in this study we have a strong magnetic field, an oscillating flow and a curved boundary. In spite of these departures from the base model, the self-similar evolution can still be clearly distinguished. This implies that this is a robust physical mechanism that could occur in flows in many systems with strong magnetic fields. As the self-similar model comes from dimensional analysis of the system, it inherently has unknown constants associated with it. For the case presented in this paper, the constant we determined was consistent with previous hydrodynamic studies of shear layer mixing. However, a wider parameter survey is necessary to determine if this constant has some as yet unknown dependency on the system that is as yet not built into the model.

In the second stage of the evolution, as the energy of the turbulent motions dominates the energy of the coherent wave motions the model we propose to understand this phase focuses on the decay of this turbulent energy. This model provides a good prediction of the temporal decay of the turbulent energy of the simulation. This allows for a reasonable prediction of the heating in the simulation due to this turbulent decay to be estimated. One key aspect of this model is we assume that the characteristic length-scale of the turbulence remains fixed. However, the growth of the turbulent area shown by the difference {between} the blue and red lines in Figure \ref{tube_area} implies that even after $t\approx 50$ the area does increase. This could imply that there is also some growth in the largest scale of the turbulence in the layer over time. This would change the exponent of the power law of the turbulent decay \citep[e.g.][]{K41, Oberlack2002}. Clearly, the simulated energy evolves at close to $\sim t^{-2}$, but it would be interesting to explore how including an evolving length-scale improves the modelling of the system.

With both phases of dynamics, what is striking about the models proposed is that they both are taken almost directly from their hydrodynamic equivalents without major revision due to the presence of strong magnetic fields. We have implicitly used the magnetic fields in both phases, mainly through assuming that Alfv\'{e}n waves are travelling back and forth along the magnetic field to connect turbulent fluctuations in one region in $z$ to other regions in $z$. In the first phase this implies the influence of the mixing at the apex is felt the full length of the tube, creating velocity and magnetic field fluctuations all along the tube, and allowing the coherent oscillations of the mixing layer to develop. For the second phase, this connection means that we assume turbulence has developed along the full length of the tube and assuming the magnitude of the turbulent energy (whether it is in turbulent kinetic energy or turbulent magnetic energy) is the same along the tube due to the magnetic field connecting regions along $z$. This assumption allows us to calculate heating rates for the system.

Another consequence of the magnetic field is it makes the turbulent motions highly anisotropic (with KHi swirls being highly elongated along the $z$-direction \citep[e.g.][]{antolin18} giving quasi-2D like behaviour. This may imply that the value of the constant $C_1$ should be nearer to $0.5$ than the value of $\sim0.3$ we have found for the first phase of the dynamics. However, we know that some of the energy of the turbulence is held in magnetic fluctuations, which will reduce the magnitude of the turbulent motions, slowing mixing. Assuming that there is equipartition between the energy of velocity and magnetic field fluctuations this would correspond to a factor of $1/\sqrt{2}$ in Equation \ref{final_h_eqn} which would increase the value found for $C_1$ by $\sqrt{2}$. Taking that the velocity shear is maximal at the apex and zero at $z=0$ because of the nature of the MHD kink wave under study, this would again put another factor of $1/\sqrt{2}$ in Equation \ref{final_h_eqn} and another subsequent increase in the value found for $C_1$ by $\sqrt{2}$. 
Therefore, in some sense part of the effect of the magnetic field on the turbulence model is wrapped up in the value of the constant $C_1$. 

\subsection{Resonant absorption and its lack of consequence in our simulations}

For the simulations presented in this paper, as with many others of KHi turbulence developing from oscillations of tubes with thin boundaries \citep{TERR2008,MAGYAR2016},  the turbulent mixing broadens the tube boundary. This takes the tube density profile to be one where resonant layers \citep{goossens92} would exist for a linear wave perturbation. Therefore, it may be expected that resonant absorption could develop in the tube once mixing has made a clear boundary layer between the inside and outside of the tube. However, {we find no evidence} of a resonant layer forming in the mixing layer at any point in our simulation and to explain the evolution of the turbulence in our simulation our model does not need to invoke resonant absorption at all. {Beyond the visual lack of a resonant layer, both panels of Figure \ref{core_mom_damping} show that the momentum and energy of the core of the tube are accurately accounted for by the model we propose. This can only be the case if the only change in the momentum and kinetic energy of the core is because the turbulent layer is eroding the core. If there where significant energy transfer from the core to the external layer through resonant absorption, these predictions would overestimate the momentum and kinetic energy of the core. Though we cannot rule out local field lines in the mixing layer resonantly absorbing some energy from the core of the tube, it means that if it is there this process is dynamically unimportant.}

So why is resonant absorption not present {(or if it is, dynamically unimportant)}? The simple answer is that the development of turbulence naturally inhibits the formation of a resonant layer. There are two explanations as to why we should not expect it to develop. Firstly, the KHi turbulence is constantly driving changes in the local conditions of the mixing layer meaning that the resonant field lines are constantly changing and through the swirling motion of the turbulence field lines with different Alfv\'{e}n speed are locked together and unable to freely move (which inhibits them developing their own oscillations). Secondly (or another way of looking at this) is that turbulence at small scales can act to have a diffusion effect at larger scales \citep[e.g.][]{Taylor_1921}. We can infer that there would be a strongly enhanced magnetic diffusion felt by the tube in the mixing layer as a consequence of turbulent motions {\citep[c.f. turbulent magnetic diffusion in mean-field dynamo models][]{BRAN2023}.} As the efficacy of resonant absorption is greatly reduced in the presence of strong diffusion, it can be understood that turbulence would work to suppress the development of a resonant layer. It is worth noting that the magnitude of the turbulent diffusion scales with the magnitude of the velocity shear. Therefore, smaller shear flows will produce less turbulent diffusion and at sufficiently small magnitude may allow the resonant absorption process to manifest.

{To understand which will be dominate, resonant absorption or the Kelvin-Helmholtz instability, we can consider the timescales of the system. In the initial evolution, if the KHi timescale is much shorter than that of resonant absorption, e.g. with thin boundary layers or large velocity amplitudes of the waves, then the KHi will likely grow and at no time will resonant absorption be important. However, in the converse situation, e.g. with large boundary layers or small amplitudes of the waves, then resonant absorption will dominate leading to the development of a resonant layer. However, this resonant layer is KHi unstable \citep[e.g.][]{TERR2008, ANT2015} which will drive turbulence that can kill the resonant absorption process.}

There is, however, one further point to consider. {The oscillation of the mixing layer (where a resonant layer could develop) appears to no longer be that of the original kink oscillation. This is its own oscillation being forced by momentum injection into the layer by mixing. Its frequency is not the kink frequency of the system but a hybrid of that frequency and the kink frequency of the layer itself. Therefore it is likely any resonant process would be associated with this new oscillation of the mixing layer and not that of the initial kink oscillation. If this is the case, then the growth of the KHi would completely remove the possibility of resonant absorption of the global kink mode from the system, only allowing energy to be absorbed from large-scale motions of the mixing layer to local resonances, which would preclude resonant absorption as a damping mechanism once the KHi mixing layer has developed.} 

\subsection{Why is this model different from the other ``KHi'' damping model}

The model presented by \citet{VD2021} is labelled in their work as damping by the Kelvin-Helmholtz instability. However, the model presented in this paper and their model have vastly different formulations, with very different dependency on the density contrast (and in the case of the model in this paper two distinct regimes). This leads to the question: How can two different formulations be modelling the same physical phenomenon? The model presented in this paper is developed from direct analysis of the turbulence created by the Kelvin-Helmholtz instability and, as shown by the comparison in \citet{AH2020} and Section \ref{sim_compare}, gives results that are consistent with those found in simulations of loop oscillations that develop the Kelvin-Helmholtz instability. 

The model presented in \citet{VD2021} used the linear eigenfunction for a kink wave and then prescribed a nonlinear amplitude to the oscillation. By integrating the wave oscillation over one period they could calculate the energy extraction rate through these nonlinearities. As they only study the particular wave mode associated with the kink wave, there are no other modes that are excited in the system. This means that there is no perturbation to seed the linear growth of the Kelvin-Helmholtz instability, and as a consequence there can be no Kelvin-Helmholtz turbulence. That is to say, the model they calculated does not allow the growth of the Kelvin-Helmholtz instability so cannot have any damping through Kelvin-Helmholtz turbulence.

This leads to the question, what is the nature of the nonlinear damping investigated in \citet{VD2021}. The initial conditions of their model are not an exact solution of the nonlinear equations (unlike a linear Alfv\'{e}n wave in incompressible nonlinear MHD). Therefore, the kink wave (characterised by a wavenumber $k_0$ and azimuthal wavemode $m=1$) does work on the neighbouring Fourier mode (in this case with wavenumber $k=2k_0$ and azimuthal wave number $m=2$). As with any forcing problem, the energy will be trapped in that wave mode efficiently if the forcing (which occurs at twice the kink frequency), and the wave mode being forced are resonant \citep[this is the basis of any wave turbulence argument, e.g.][]{GOLDREICH1995}. This clearly is a different mechanism to KHi damping of a kink wave as presented in this paper, highlighting that when considering nonlinear mechanisms for wave damping there are many different ways these nonlinearities can manifest.

\subsection{Heating rates and the implications for coronal heating}

One of the important predictions of the models present here are the heating rates from both stages of the evolution. To make this comparison, we need to compare the characteristic heating rate of the simulation with the cooling time for comparable coronal plasma. The cooling time for the solar corona (based on a number density of $10^9$\,cm$^{-3}$) is $\tau_{\rm cool}\sim 10^3$\,s. In the time units of our simulation (i.e. sound crossing times), if we assume a loop length of $10^{10}$\,cm and a sound speed of $1.2\times 10^7$\,cm\,s$^{-1}$ the dimensional cooling time becomes $24\tau$. As this is an exponential decay, it is the equivalent of the thermal energy reducing by a factor of $1/\exp(1)$. Note this ignores the losses through thermal conduction to the chromosphere and so should be taken merely as a crudely representative value.

For the heating, we can see from Figure \ref{heating}, at $t\approx 40$ the internal energy had increased by a value of $\sim 0.018$. This gives a rate of increase of internal energy of $0.00045$. By dividing the total internal energy at the start of the simulation divided by $\exp(1)$ by this number we get a heating time of $\sim 10^4\tau$, which is significantly larger than the cooling time of $24\tau$. Therefore, even though the kink wave in this simulation is a relatively nonlinear perturbation (with the initial kick of 20\% of the sound speed), the heating rates found are significantly smaller than those needed to balance radiative losses. The small heating rates we find in our simulation are consistent with those found by \citet{HOW2017} where explicit viscosity and magnetic resistivity were included.

\subsection{A corollary on coronal seismology}

The work presented in this paper leads to a very important corollary about the damping of kink waves in the solar atmosphere. It is standard to fit either exponential or, in some cases, Gaussian damping envelopes to the observed decay of loop oscillations \citep[e.g.][]{NEC2019}. However, these damping envelopes are fundamentally connected to linear wave theory.

The work presented in Sections \ref{MOM_section} and \ref{AMP_pred} highlights how KHi turbulence produces non-linear wave damping. The wave damping envelope for this nonlinear mechanism has a different functional form to the linear damping profiles. Even more importantly, exactly how the tube motions are measured changes the damping envelope that is measured. This implies that when measuring damping of kink waves from observations of the solar corona, it is important to consider more than just linear damping envelopes to understand the evolution.

\begin{acknowledgments}
AH is supported by STFC Research Grant No. ST/V000659/1.
IA and AH are supported by project PID2021-127487NB-I00 from Ministerio de Ciencia e Innovaci\'on and FEDER funds.
AH and TM are supported by JSPS KAKENHI Grant Number 19K03669.
AH would like to acknowledge the discussions with members of ISSI Team 457 ``The Role of Partial Ionization in the Formation, Dynamics and Stability of Solar Prominences'', which have helped improve the ideas in this manuscript.
AH and TM would like to acknowledge support by the Research Institute for Mathematical Sciences, an International Joint Usage/Research Center located in Kyoto University. 
For the purpose of open access, the author has applied a ‘Creative Commons Attribution (CC BY) licence to any Author Accepted Manuscript version arising.\\
\\
The active branch of the (P\underline{I}P) code is available on GitHub (\url{https://github.com/AstroSnow/PIP}). {Data required to reproduce the figures is freely available to download from Zenodo \href{10.5281/zenodo.10655009}{doi:10.5281/zenodo.10655009}}. All simulation data is available upon reasonable request.
\end{acknowledgments}

\bibliographystyle{aasjournal}
\bibliography{NLKH}

\begin{thebibliography}{}
\expandafter\ifx\csname natexlab\endcsname\relax\def\natexlab#1{#1}\fi
\providecommand{\url}[1]{\href{#1}{#1}}
\providecommand{\dodoi}[1]{doi:~\href{http://doi.org/#1}{\nolinkurl{#1}}}
\providecommand{\doeprint}[1]{\href{http://ascl.net/#1}{\nolinkurl{http://ascl.net/#1}}}
\providecommand{\doarXiv}[1]{\href{https://arxiv.org/abs/#1}{\nolinkurl{https://arxiv.org/abs/#1}}}

\bibitem[{{Antolin} {et~al.}(2015){Antolin}, {Okamoto}, {De Pontieu},
  {Uitenbroek}, {Van Doorsselaere}, \& {Yokoyama}}]{ANT2015}
{Antolin}, P., {Okamoto}, T.~J., {De Pontieu}, B., {et~al.} 2015, \apj, 809,
  72, \dodoi{10.1088/0004-637X/809/1/72}

\bibitem[{{Antolin} {et~al.}(2018){Antolin}, {Pagano}, {De Moortel}, \&
  {Nakariakov}}]{antolin18}
{Antolin}, P., {Pagano}, P., {De Moortel}, I., \& {Nakariakov}, V.~M. 2018,
  \apjl, 861, L15, \dodoi{10.3847/2041-8213/aacf98}

\bibitem[{{Antolin} {et~al.}(2014){Antolin}, {Yokoyama}, \& {Van
  Doorsselaere}}]{ANT2014}
{Antolin}, P., {Yokoyama}, T., \& {Van Doorsselaere}, T. 2014, \apj, 787, L22,
  \dodoi{10.1088/2041-8205/787/2/L22}

\bibitem[{{Arregui}(2021)}]{ARREGUI2021}
{Arregui}, I. 2021, \apjl, 915, L25, \dodoi{10.3847/2041-8213/ac0d53}

\bibitem[{Aschwanden {et~al.}(1999)Aschwanden, Fletcher, Schrijver, \&
  Alexander}]{aschwanden99}
Aschwanden, M.~J., Fletcher, L., Schrijver, C.~J., \& Alexander, D. 1999, The
  Astrophysical Journal, 520, 880, \dodoi{10.1086/307502}

\bibitem[{{Baltzer} \& {Livescu}(2020)}]{BALTZER2020}
{Baltzer}, J.~R., \& {Livescu}, D. 2020, Journal of Fluid Mechanics, 900, A16,
  \dodoi{10.1017/jfm.2020.466}

\bibitem[{{Brandenburg} {et~al.}(2023){Brandenburg}, {Elstner}, {Masada}, \&
  {Pipin}}]{BRAN2023}
{Brandenburg}, A., {Elstner}, D., {Masada}, Y., \& {Pipin}, V. 2023, \ssr, 219,
  55, \dodoi{10.1007/s11214-023-00999-3}

\bibitem[{{Brown} \& {Roshko}(1974)}]{BROWN1974}
{Brown}, G.~L., \& {Roshko}, A. 1974, Journal of Fluid Mechanics, 64, 775,
  \dodoi{10.1017/S002211207400190X}

\bibitem[{{Chen} \& {Schuck}(2007)}]{CHEN2007}
{Chen}, J., \& {Schuck}, P.~W. 2007, \solphys, 246, 145,
  \dodoi{10.1007/s11207-007-9011-9}

\bibitem[{{Goddard} \& {Nakariakov}(2016)}]{goddard16b}
{Goddard}, C.~R., \& {Nakariakov}, V.~M. 2016, A\&A, 590, L5,
  \dodoi{10.1051/0004-6361/201628718}

\bibitem[{{Goldreich} \& {Sridhar}(1995)}]{GOLDREICH1995}
{Goldreich}, P., \& {Sridhar}, S. 1995, \apj, 438, 763, \dodoi{10.1086/175121}

\bibitem[{{Goossens} {et~al.}(2006){Goossens}, {Andries}, \&
  {Arregui}}]{goossens06}
{Goossens}, M., {Andries}, J., \& {Arregui}, I. 2006, Philosophical
  Transactions of the Royal Society of London Series A, 364, 433,
  \dodoi{10.1098/rsta.2005.1708}

\bibitem[{{Goossens} {et~al.}(2002){Goossens}, {Andries}, \&
  {Aschwanden}}]{goossens02a}
{Goossens}, M., {Andries}, J., \& {Aschwanden}, M.~J. 2002, \aap, 394, L39,
  \dodoi{10.1051/0004-6361:20021378}

\bibitem[{{Goossens} {et~al.}(2011){Goossens}, {Erd{\'e}lyi}, \&
  {Ruderman}}]{goossens11}
{Goossens}, M., {Erd{\'e}lyi}, R., \& {Ruderman}, M.~S. 2011, \ssr, 158, 289,
  \dodoi{10.1007/s11214-010-9702-7}

\bibitem[{{Goossens} {et~al.}(1992){Goossens}, {Hollweg}, \&
  {Sakurai}}]{goossens92}
{Goossens}, M., {Hollweg}, J.~V., \& {Sakurai}, T. 1992, \solphys, 138, 233,
  \dodoi{10.1007/BF00151914}

\bibitem[{{Hasegawa} {et~al.}(2006){Hasegawa}, {Fujimoto}, {Takagi}, {Saito},
  {Mukai}, \& {R{\`e}Me}}]{HASEGAWA2006}
{Hasegawa}, H., {Fujimoto}, M., {Takagi}, K., {et~al.} 2006, Journal of
  Geophysical Research (Space Physics), 111, A09203,
  \dodoi{10.1029/2006JA011728}

\bibitem[{{Hillier}(2019)}]{HILL2019b}
{Hillier}, A. 2019, Physics of Plasmas, 26, 082902, \dodoi{10.1063/1.5103248}

\bibitem[{{Hillier} \& {Arregui}(2019)}]{HILL2019c}
{Hillier}, A., \& {Arregui}, I. 2019, \apj, 885, 101,
  \dodoi{10.3847/1538-4357/ab4795}

\bibitem[{{Hillier} {et~al.}(2019){Hillier}, {Barker}, {Arregui}, \&
  {Latter}}]{HILL2019}
{Hillier}, A., {Barker}, A., {Arregui}, I., \& {Latter}, H. 2019, \mnras, 482,
  1143, \dodoi{10.1093/mnras/sty2742}

\bibitem[{{Hillier} {et~al.}(2023){Hillier}, {Snow}, \&
  {Arregui}}]{HILLIER2023}
{Hillier}, A., {Snow}, B., \& {Arregui}, I. 2023, \mnras, 520, 1738,
  \dodoi{10.1093/mnras/stad234}

\bibitem[{{Hillier} {et~al.}(2016){Hillier}, {Takasao}, \&
  {Nakamura}}]{HILL2016}
{Hillier}, A., {Takasao}, S., \& {Nakamura}, N. 2016, \aap, 591, A112,
  \dodoi{10.1051/0004-6361/201628215}

\bibitem[{{Hillier} {et~al.}(2020){Hillier}, {Van Doorsselaere}, \&
  {Karampelas}}]{AH2020}
{Hillier}, A., {Van Doorsselaere}, T., \& {Karampelas}, K. 2020, \apjl, 897,
  L13, \dodoi{10.3847/2041-8213/ab9ca3}

\bibitem[{{Howson} {et~al.}(2017){Howson}, {De Moortel}, \&
  {Antolin}}]{HOW2017}
{Howson}, T.~A., {De Moortel}, I., \& {Antolin}, P. 2017, \aap, 607, A77,
  \dodoi{10.1051/0004-6361/201731178}

\bibitem[{{Kolmogorov}(1941)}]{K41}
{Kolmogorov}, A. 1941, Akademiia Nauk SSSR Doklady, 30, 301

\bibitem[{{Lin} {et~al.}(2007){Lin}, {Engvold}, {Rouppe van der Voort}, \& {van
  Noort}}]{lin07}
{Lin}, Y., {Engvold}, O., {Rouppe van der Voort}, L.~H.~M., \& {van Noort}, M.
  2007, \solphys, 246, 65, \dodoi{10.1007/s11207-007-0402-8}

\bibitem[{{Magyar} \& {Van Doorsselaere}(2016)}]{MAGYAR2016}
{Magyar}, N., \& {Van Doorsselaere}, T. 2016, \aap, 595, A81,
  \dodoi{10.1051/0004-6361/201629010}

\bibitem[{{McIntosh} {et~al.}(2011){McIntosh}, {de Pontieu}, {Carlsson},
  {Hansteen}, {Boerner}, \& {Goossens}}]{mcintosh11}
{McIntosh}, S.~W., {de Pontieu}, B., {Carlsson}, M., {et~al.} 2011, \nat, 475,
  477, \dodoi{10.1038/nature10235}

\bibitem[{Nakariakov {et~al.}(1999)Nakariakov, Ofman, DeLuca, Roberts, \&
  Davila}]{nakariakov99}
Nakariakov, V.~M., Ofman, L., DeLuca, E.~E., Roberts, B., \& Davila, J.~M.
  1999, Science, 285, 862, \dodoi{10.1126/science.285.5429.862}

\bibitem[{{Nakariakov} {et~al.}(2021){Nakariakov}, {Anfinogentov}, {Antolin},
  {Jain}, {Kolotkov}, {Kupriyanova}, {Li}, {Magyar}, {Nistic{\`o}}, {Pascoe},
  {Srivastava}, {Terradas}, {Vasheghani Farahani}, {Verth}, {Yuan}, \&
  {Zimovets}}]{nakariakov21}
{Nakariakov}, V.~M., {Anfinogentov}, S.~A., {Antolin}, P., {et~al.} 2021, \ssr,
  217, 73, \dodoi{10.1007/s11214-021-00847-2}

\bibitem[{{Nechaeva} {et~al.}(2019){Nechaeva}, {Zimovets}, {Nakariakov}, \&
  {Goddard}}]{NEC2019}
{Nechaeva}, A., {Zimovets}, I.~V., {Nakariakov}, V.~M., \& {Goddard}, C.~R.
  2019, \apjs, 241, 31, \dodoi{10.3847/1538-4365/ab0e86}

\bibitem[{Oberlack(2002)}]{Oberlack2002}
Oberlack, M. 2002, PAMM, 1, 294,
  \dodoi{https://doi.org/10.1002/1617-7061(200203)1:1<294::AID-PAMM294>3.0.CO;2-W}

\bibitem[{{Onsager}(1949)}]{ONSAGER1949}
{Onsager}, L. 1949, Il Nuovo Cimento, 6, 279, \dodoi{10.1007/BF02780991}

\bibitem[{{Roberts}(1983)}]{roberts83}
{Roberts}, B. 1983, \solphys, 87, 77

\bibitem[{{Roberts}(2000)}]{roberts00}
---. 2000, \solphys, 193, 139

\bibitem[{{Ruderman} \& {Erd{\'e}lyi}(2009)}]{ruderman09}
{Ruderman}, M.~S., \& {Erd{\'e}lyi}, R. 2009, \ssr, 149, 199,
  \dodoi{10.1007/s11214-009-9535-4}

\bibitem[{Ruderman \& Roberts(2002)}]{ruderman02}
Ruderman, M.~S., \& Roberts, B. 2002, The Astrophysical Journal, 577, 475,
  \dodoi{10.1086/342130}

\bibitem[{{Sagaut} \& {Cambon}(2018)}]{Sagaut2018}
{Sagaut}, P., \& {Cambon}, C. 2018, {Homogeneous Turbulence Dynamics},
  \dodoi{10.1007/978-3-319-73162-9}

\bibitem[{{Sakurai} {et~al.}(1991){Sakurai}, {Goossens}, \&
  {Hollweg}}]{sakurai91}
{Sakurai}, T., {Goossens}, M., \& {Hollweg}, J.~V. 1991, \solphys, 133, 227,
  \dodoi{10.1007/BF00149888}

\bibitem[{{Schmieder} {et~al.}(2013){Schmieder}, {Kucera}, {Knizhnik}, {Luna},
  {Lopez-Ariste}, \& {Toot}}]{schmieder13}
{Schmieder}, B., {Kucera}, T.~A., {Knizhnik}, K., {et~al.} 2013, \apj, 777,
  108, \dodoi{10.1088/0004-637X/777/2/108}

\bibitem[{Taylor(1922)}]{Taylor_1921}
Taylor, G.~I. 1922, Proceedings of the London Mathematical Society, s2-20, 196,
  \dodoi{https://doi.org/10.1112/plms/s2-20.1.196}

\bibitem[{{Taylor}(1935)}]{Taylor1935}
{Taylor}, G.~I. 1935, Proceedings of the Royal Society of London Series A, 151,
  421, \dodoi{10.1098/rspa.1935.0158}

\bibitem[{{Terradas} {et~al.}(2008){Terradas}, {Andries}, {Goossens},
  {Arregui}, {Oliver}, \& {Ballester}}]{TERR2008}
{Terradas}, J., {Andries}, J., {Goossens}, M., {et~al.} 2008, \apj, 687, L115,
  \dodoi{10.1086/593203}

\bibitem[{{Terradas} \& {Ofman}(2004)}]{TERRADAS2004}
{Terradas}, J., \& {Ofman}, L. 2004, \apj, 610, 523, \dodoi{10.1086/421514}

\bibitem[{{Tomczyk} {et~al.}(2007){Tomczyk}, {McIntosh}, {Keil}, {Judge},
  {Schad}, {Seeley}, \& {Edmondson}}]{tomczyk07}
{Tomczyk}, S., {McIntosh}, S.~W., {Keil}, S.~L., {et~al.} 2007, Science, 317,
  1192, \dodoi{10.1126/science.1143304}

\bibitem[{{van Ballegooijen} {et~al.}(2011){van Ballegooijen}, {Asgari-Targhi},
  {Cranmer}, \& {DeLuca}}]{VANBALLEGOOIJEN2011}
{van Ballegooijen}, A.~A., {Asgari-Targhi}, M., {Cranmer}, S.~R., \& {DeLuca},
  E.~E. 2011, \apj, 736, 3, \dodoi{10.1088/0004-637X/736/1/3}

\bibitem[{{Van Doorsselaere} {et~al.}(2004){Van Doorsselaere}, {Andries},
  {Poedts}, \& {Goossens}}]{VanDoorsselaere2004}
{Van Doorsselaere}, T., {Andries}, J., {Poedts}, S., \& {Goossens}, M. 2004,
  \apj, 606, 1223, \dodoi{10.1086/383191}

\bibitem[{{Van Doorsselaere} {et~al.}(2021){Van Doorsselaere}, {Goossens},
  {Magyar}, {Ruderman}, \& {Ismayilli}}]{VD2021}
{Van Doorsselaere}, T., {Goossens}, M., {Magyar}, N., {Ruderman}, M.~S., \&
  {Ismayilli}, R. 2021, \apj, 910, 58, \dodoi{10.3847/1538-4357/abe630}

\bibitem[{Winant \& Browand(1974)}]{winant_browand_1974}
Winant, C.~D., \& Browand, F.~K. 1974, Journal of Fluid Mechanics, 63,
  237–255, \dodoi{10.1017/S0022112074001121}

\end{thebibliography}

\end{document}